\documentclass[12pt]{article}
\usepackage{amssymb,amscd,array}
\catcode `\@=11
\@addtoreset{equation}{section}

\def\harr#1#2{\smash{\mathop{\hbox to .5in{\rightarrowfill}}
\limits^{\scriptstyle#1}_{\scriptstyle#2}}}

\def\harrl#1#2{\smash{\mathop{\hbox to .5in{\leftarrowfill}}
\limits^{\scriptstyle#1}_{\scriptstyle#2}}}

\def\qed{\hspace*{\fill}\rule{3mm}{3mm}}

\newcommand{\be}{\begin{equation}}
\newcommand{\ee}{\end{equation}}
\newcommand{\bea}{\begin{eqnarray}}
\newcommand{\eea}{\end{eqnarray}}

\newcommand{\R}{\mathbb{R}}

\newcommand{\C}{\mathbb{C}}
\newtheorem{thm}{Theorem}[section]
\newtheorem{rem}[thm]{Remark}
\newtheorem{lemma}[thm]{Lemma}

\newtheorem{prop}[thm]{Proposition}
\begin{document}
\begin{titlepage}
\begin{center}
{\bf \Large{The Quantum Supersymmetric Vector Multiplet \\
and Some Problems in Non-Abelian Supergauge Theory \\}}
\end{center}
\vskip 1.0truecm
\centerline{Dan Radu Grigore
\footnote{e-mail: grigore@physik.unizh.ch}
\footnote{Permanent address: Dept. Theor. Phys., Inst. Atomic Phys.,
Bucharest-M\u agurele, MG 6, Rom\^ania}
and G\"unter Scharf
\footnote{e-mail: scharf@physik.unizh.ch}}
\vskip5mm
\centerline{Institute of Theor. Phys., University of Z\"urich}
\centerline{Winterthurerstr., 190, Z\"urich CH-8057, Switzerland}
\vskip 2cm
\bigskip \nopagebreak
\begin{abstract}
\noindent
We consider the supersymmetric vector multiplet in a purely quantum framework.
We obtain some discrepancies with respect to the literature in the expression
of the super-propagator and we prove that the model is consistent
only for positive mass. The gauge structure is constructed purely deductive and
leads to the necessity of introducing scalar ghost superfields, in analogy
to the usual gauge theories. The construction of a
consistent supersymmetric
gauge theory based on the vector
 model depends crucially one the definition
of gauge invariance. We find some significant difficulties to impose a
supersymmetric gauge invariance condition for the usual expressions from the
literature.
\end{abstract}
PACS: 11.10.-z, 11.30.Pb

\newpage\setcounter{page}1
\end{titlepage}
\section{Introduction}

The supersymmetric gauge theories are constructed using the so-called
vector supersymmetric multiplet \cite{FP} (see also \cite{WB}, \cite{Wes}
\cite{Wei}, \cite{GGRS}, \cite{Bi}, 
\cite{Fi}, \cite{Li}, \cite{So},
\cite{Pi1}, \cite{Pi2}, \cite{Sa}, 
 etc.) The justification for this
choice comes from the analysis of the unitary irreducible representations
of the
$N = 1$
supersymmetric extension of the Poincar\'e group; there are two irreducible
massive representations
\bea
\Omega_{1/2} \sim [m,0] \oplus [m,1/2] \oplus [m,1/2] \oplus [m,1],
\nonumber \\
\Omega_{1} \sim [m,1/2] \oplus [m,1] \oplus [m,1] \oplus [m,3/2],
\label{decomp}
\eea
containing a spin 1 system (see for instance \cite{RY1}; here
$[m,s]$
is the irreducible representation of mass $m$ and spin $s$ of the Poincar\'e
group.) The standard vector multiplet is 
constructed such that the one-particle
subspace of the Fock space carries the ``simplest" representation
$\Omega_{1/2}$.
We will prove in this paper that it is very hard to build a corresponding
fully consistent quantum theory with all the usual properties. The other
possibility 
is to
 construct a supersymmetric multiplet for which the associated
Fock space 
has
$\Omega_{1}$
as the one-particle subspace of the Fock space. The construction in this
case is natural and straightforward; 
 the content of this multiplet is
a (complex) spin
 1 and a spin 3/2 fields (more precisely a
Rarita-Schwinger field without the transversality conditions.) We have
proved in \cite{GS1} that the second multiplet can be the
basis for a 
supersymmetric extension of quantum gauge theory because
its gauge structure involving
 ghosts, anti-ghosts and
 unphysical scalar
(Goldstone) fields is very similar to the ordinary
 gauge theory.

In this paper we consider in detail the
$\Omega_{1/2}$
vector model (so from no on, when we say ``the vector model" we mean the
the
$\Omega_{1/2}$
vector model.) We intend to give a rigorous treatment of all aspects of this
model using the Epstein-Glaser framework. This seems to be a rather difficult
task. In this paper we start this program analysing the layout of the model,
that is the construction of the quantum multiplet, its gauge structure and
the expression of the interaction Lagrangian (or, in the language of
perturbation theory, the first order chronological product).
The analysis will be performed entirely in the quantum framework \cite{Lo},
\cite{O}, \cite{NK}, \cite{CGS}, \cite{Gr2}, \cite{GS1} avoiding the
usual approach based on  quantizing a classical supersymmetric  theory.
In this way
 we avoid the complications associated  to the proper
mathematical definition of a super-manifold
\cite{Fr}, \cite{IAS} and we
 do not need a quantization procedure.

The main results are the following. First we show that the vector model
is consistent only for positive mass. Next, we determine the gauge
structure of the vector model: it coincides essentially with the
expression from the literature but, because the mass of the multiplet is
positive we need to introduce some scalar ghost superfields. We are also
able to determine the general expressions for the Feynman super-propagators
in a purely deductive way; some discrepancy with the standard literature
appears. Finally, we investigate the  expression for the interaction Lagrangian
consistent with the conventional approaches based on path integral quantization.

It is important to outline the mathematical framework used in this paper
from the very beginning. The description of higher spin fields will be
done in the indefinite metric approach (Gupta-Bleuler). That is, we
construct a Hilbert space
${\cal H}$
with a non-degenerate
sesqui-linear form
$<\cdot, \cdot>$
and a gauge charge operator $Q$ verifying
$Q^{2} = 0$;
the form
$<\cdot, \cdot>$
becomes positively defined when restricted to a factor Hilbert space
${\cal H}_{phys} \equiv Ker(Q)/Im(Q)$
which will be the physical space of our problem. 
The interaction Lagrangian
$t(x)$
will be some Wick polynomial acting in the total Hilbert space
${\cal H}$
and verifying the conditions
\bea
[Q, t(x)]= i\partial_{\mu} t^{\mu}(x)
\label{gauge}
\eea
for some Wick polynomials
$t^{\mu}(x)
$;
this condition guarantees that the interaction Lagrangian
$t(x)$
factorises to the physical Hilbert space
$Ker(Q)/Im(Q)$
in the adiabatic limit, i.e. after integration over $x$; the condition
(\ref{gauge}) is equivalent to the usual condition of (free) current
conservation.
 The condition (\ref{gauge}) has far reaching physical
consequences:under some
 reasonable additional assumptions one can prove
that the usual expression
 of the interaction Lagrangian for a Yang-Mills
model is unique, up to trivial
 terms. It is desirable to generalize this
scheme to supersymmetric theories.

In the supersymmetric framework one postulates that the basic supersymmetric
multiplets should be organized in superfields i.e. fields dependent on
space-time variables and some auxiliary Grassmann parameters. It is showed
in \cite{GS1} that one can consistently replace fields by superfields: one
has a canonical map
$w \mapsto sw \equiv W$
mapping a ordinary Wick monomial
$w(x)$
into its supersymmetric extension
$W(x,\theta,\bar{\theta})$; in particular this map associates to every field
of the model a superfield. Moreover, one postulates that the interaction
Lagrangian $t$ should be of the form
\be
t(x) \equiv \int d\theta^{2} d\bar{\theta}^{2} T(x,\theta,\bar{\theta})
\ee
for some supersymmetric Wick polynomial $T$. This hypothesis makes
possible the generalization of the Epstein-Glaser approach to the
supersymmetric case as it is showed in \cite{GS1}.

Concerning the gauge invariance of the model there are two possible attitudes.
One is to impose only (\ref{gauge}); this ``minimal" possibility is certainly
consistent from the physical point of view but in this case one loses the
unicity results concerning the interaction Lagrangian. One can hope to
keep this unicity result if one finds out a supersymmetric generalization of
(\ref{gauge}). A natural candidate would be the relation:
\bea
[Q, T(x,\theta,\bar{\theta}) ] =
i\partial_{\mu} T^{\mu}(x,\theta,\bar{\theta})
+ \dots
\label{gauge-susy}
\eea
where by $\dots$ we mean total divergence expressions in the Grassmann
variables. It is clear that (\ref{gauge-susy}) implies (\ref{gauge})
but not conversely. We call (\ref{gauge-susy}) the condition of
{\it supersymmetric gauge invariance}. In \cite{GS1} we have showed that the
stronger condition (\ref{gauge-susy}) can be imposed and indeed the unicity
argument concerning the interaction Lagrangian holds. However for the
$\Omega_{1/2}$
vector model the situation is not so good. If one uses only the ``minimal"
gauge invariance condition (\ref{gauge}) then one loses the unicity of the
interaction Lagrangian. If one tries to impose the suspersymmetric version
(\ref{gauge-susy}) one finds out that the usual expressions for the
interaction Lagrangian suggested by the literature do not fulfil it. Of
course it is in principle possible to find alternative expression for the
interaction Lagrangian such that (\ref{gauge-susy}) is true, but this
possibility seems to be rather unprobable. So our results concerning the
construction of the interaction Lagrangian for the
$\Omega_{1/2}$
vector model must be considered as a criticism of the traditional approaches
based on the path integral formalism: a rigorous approach produces some
differences and negative results.

The structure of the paper is the following.
 In Section \ref{qsr}  we give a
brief but general discussion about 
supersymmetric multiplets and
the associated superfields.  In Section
 \ref{vector}  we give a detailed
description of the vector multiplet in
 a purely quantum framework. We find
out differences with respect to the
 literature in the expression of the
super-propagator. In Section
 \ref{chiral-multi} we first remained the basic
facts about the chiral
 multiplet (they will be used as scalar ghost
superfields) and then  we
 construct in analogy the ghost and the anti-ghost
superfields. In particular we prove that so-called Wess-Zumino gauge is not
a supersymmetric decomposition of the vector superfield into a chiral,
anti-chiral and a ``physical" part.
In Section
 \ref{gauge-charge} we use these superfields for the
construction of the
 gauge structure.
 In Section \ref{sm}
 we study the
interaction Lagrangian which can be inferred from the
 expression appearing
in the formal path integral quantization method 
(combined with the
Faddeev-Popov trick). We have to add to it supplementary
 terms containing
the scalar ghosts superfields and argue that it cannot
 fulfil the
supersymmetric gauge invariance condition (\ref{gauge-susy}).
 In Section
\ref{lin} we investigate the possibility of using the linear vector
multiplet in a supersymmetric gauge theory.

One can conclude that the new vector multiplet proposed for the first time in
\cite{GS1} and based only on chiral superfields is a more natural
object and it remains as a  serious candidate for a possible supersymmetric
extension of the standard model.

\section{Quantum Supersymmetric Theory \label{qsr}}

We remind here the definition of a supersymmetric
 theory in a pure quantum
context. We will not consider extended
 supersymmetries here.
 We follow closely
\cite{GS1}.

The conventions are the following: (a) we use summation over dummy indices;
(b) we raise and lower Minkowski
indices with the Minkowski pseudo-metric
$g_{\mu\nu} = g^{\mu\nu}$
with diagonal
$1,-1,-1,-1$;
(c) we raise and lower Weyl indices with the anti-symmetric
$SL(2,\C)$-invariant
tensor
$\epsilon_{ab} = - \epsilon^{ab}; \quad \epsilon_{12} = 1$
;
(d) we denote by
$\sigma^{\mu}$
the usual Pauli matrices with elements denoted by
$\sigma^{\mu}_{a\bar{b}}$
and the convention
$\sigma^{0} = {\bf 1}$;
(e) we introduce the notations:
\bea
\theta\lambda \equiv \theta^{a} \lambda_{a}, \qquad
\bar{\theta}\bar{\lambda} \equiv \bar{\theta}_{\bar{a}} \bar{\lambda}^{\bar{a}},
\nonumber \\
\theta^{2} \equiv \theta\theta, \qquad
\bar{\theta}^{2} \equiv \bar{\theta}\bar{\theta}
\nonumber \\
\theta\sigma^{\mu}\bar{\lambda} \equiv
\theta^{a} \sigma^{\mu}_{a\bar{b}} \bar{\lambda}^{\bar{b}}.
\eea

Suppose that we have a quantum theory
 of {\bf free} fields; this means that
we have the following construction:
\begin{itemize}
\item
${\cal H}$
is a Hilbert space of Fock type (associated to some one-particle Hilbert space
describing some choice of elementary particles) with the scalar product
$(\cdot, \cdot)$;
\item
$\Omega \in {\cal H}$
is a special vector called the vacuum;
\item
$U_{a,A}$
is a unitary irreducible representation of
$inSL(2,\C)$
the universal covering group of the proper orthochronous Poincar\'e group such
that
$a \in \R^{4}$
is translation in the Minkowski space and
$A \in SL(2,\C)$;
\item
$b_{j}, \quad j = 1,\dots,N_{B}$ (resp. $f_{A}, \quad A = 1,\dots,N_{F}$)
are the quantum free fields of integer (resp. half-integer) spin. We assume
that the fields are linearly independent up to equations of motion;
\item
The  equations of motion do not connect distinct fields.
\end{itemize}

In practice, one considers only particles of spin
$s \leq 2$.
For the standard vector model we will consider
 only
$s \leq 1$.
In \cite{GS1} we have considered the a more unusual case namely
$1 \leq s \leq 3/2$.
The fact that we work only with free fields is very natural from the point of
view of $S$-matrix perturbation theory in the sense of Bogoliubov \cite{BS}.
Even if 
one considers a more general case, namely a Wightman theory like in
\cite{Lo},
 one still have a Fock space structure generated by the asymptotic
fields and
 some natural assumption show that the supersymmetric structure of
the 
interacting theory is preserved for the free fields.

As we have said in the Introduction, if one considers higher-spin fields
(more precisely
$s \geq 1$),
as we do here and we have done in \cite{GS1}, it is necessary to extend
somewhat this framework: one considers in
${\cal H}$
besides the usual positive definite scalar product a non-degenerate
sesqui-linear form
$<\cdot, \cdot>$
which becomes positively defined when restricted to a factor Hilbert space
$Ker(Q)/Im(Q)$
where $Q$ is some {\it gauge charge}. We denote with
$A^{\dagger}$
the adjoint of the operator $A$ with respect to
$<\cdot, \cdot>$.

It is convenient to make the description of quantum fields more explicit.
According to the usual treatment of the standard vector model, we have to
consider that we have the following fields.
\begin{itemize}
\item
A set of Bosonic scalar fields
$b^{(j)}$
of mass
$m_{j}, \quad j = 1,\dots,s$
respectively which can be taken Hermitian without losing generality. This
means that we have the following relations:
\bea
(b^{(j)})^{\dagger} = b^{(j)}, \quad j = 1,\dots,s
\label{h-1}
\eea
\bea
(\partial^{2} + m_{j}^{2}) b^{(j)} = 0, \quad j = 1,\dots,s
\label{kg-1}
\eea
\bea
\left[ b^{(j)}(x), b^{(k)}(y) \right] = - i~\delta_{jk}~ D_{m_{j}}(x-y),
\label{CCR-1}
\eea
where
$D_{m}$
is the Pauli-Jordan causal distribution of mass $m$.
\item
A Bosonic real vector field
$b_{\mu}$
of mass $m$ verifying
\bea
(b_{\mu})^{\dagger} = b_{\mu}
\label{h-2}
\eea
\bea
(\partial^{2} + m^{2}) b_{\mu} = 0
\label{kg-2}
\eea
\bea
\left[ b_{\mu}(x), b_{\rho}(y) \right] = i~g_{\mu\rho}~ D_{m}(x-y).
\label{CCR-2}
\eea
\item
A set of Fermi fields
$f^{(A)}_{a}$
of spin $1/2$ and of mass
$M_{A}, \quad A = 1,\dots,f$
which can be taken without losing generality to be Majorana fields; here
$a = 1,2$
is a Weyl index so, the transformation property of the fields with respect
to
 the group
$SL(2,\C)$
is given by the representation
$(1/2,0)$.
We define
\bea
\bar{f}^{(A)}_{\bar{a}} \equiv
(f^{(A)}_{a})^{\dagger}
\quad A = 1,\dots,f, \quad a = 1,2
\label{h-3}
\eea
i.e. the bared indices correspond to the representation
$(0,1/2)$
of the
 group
$SL(2,\C)$.
We also suppose that these fields obey Dirac equation:
\bea
i~\sigma^{\mu}_{a\bar{b}} \partial_{\mu} \bar{f}^{(A)\bar{b}}
= M_{A} f^{(A)}_{a}, \qquad
- i~\sigma^{\mu}_{a\bar{b}} \partial_{\mu} f^{(A)a}
= M_{A} \bar{f}^{(A)}_{\bar{b}}, \quad A = 1,\dots,f
\label{dirac}
\eea
and the usual causal anticommutation relation:
\bea
\left\{ f^{(A)}_{a}(x), f^{(B)}_{b}(y) \right\}
= i~\delta_{AB}~\epsilon_{ab} M_{A} D_{M_{A}}(x-y),
\nonumber \\
\left\{ f^{(A)}_{a}(x), \bar{f}^{(B)}_{\bar{b}}(y) \right\} =
\delta_{AB}~\sigma^{\mu}_{a\bar{b}}~\partial_{\mu}D_{M_{A}}(x-y).
\label{CCR-3}
\eea
\end{itemize}

All these fields have {\it bona fid\ae} representations in Fock spaces; they
can be found in standard literature.
It is appropriate to clarify now the connection between Majorana and Dirac
fields.
\begin{prop}
(i) Suppose that the Weyl spinor
$f_{a}$
verifies only Klein-Gordon equation:
\be
(\partial^{2} + m^{2}) f_{a} = 0
\ee
of positive mass $m$ and the causal anticommutation relations:
\bea
\left\{ f_{a}(x), f_{b}(y) \right\}  = 0,
\nonumber \\
\left\{ f_{a}(x), \bar{f}_{\bar{b}}(y) \right\} =
{1\over 2m^{2}}~\sigma^{\mu}_{a\bar{b}}~\partial_{\mu}D_{m}(x-y).
\label{CCR-d}
\eea

Then the bi-spinor
\bea
\psi \equiv
\left( \begin{array}{c} f_{a} \\ \bar{g}^{\bar{b}} \end{array} \right)
\eea
where
\be
\bar{g}_{\bar{b}} \equiv
- {i\over m} \sigma^{\mu}_{a\bar{b}} \partial_{\mu} f^{a}
\ee
verifies Dirac equation.
\be
(i~\gamma\cdot\partial + m) \psi = 0;
\ee
here
$\gamma^{\mu}$
are the Dirac matrices (in the chiral representation). Conversely, if
$\psi$
is a Dirac bi-spinor, its upper component
$f_{a}$
is restricted only by the Klein-Gordon equation and its lower component
is determined as above.

(ii) Let us consider the Weyl spinor
$f_{a}$
verifying Klein-Gordon equation and let us define the following Majorana spinors:
\be
\xi^{(1)}_{a} = i~mf_{a} +
\sigma^{\mu}_{a\bar{b}}~\partial_{\mu}\bar{f}^{\bar{b}}
\qquad
\xi^{(2)}_{a} = mf_{a} +
i~\sigma^{\mu}_{a\bar{b}}~\partial_{\mu}\bar{f}^{\bar{b}}.
\ee

Then the spinors
$\xi^{(j)}_{a}, \quad j = 1,2$
verify the Dirac equation:
\bea
i~\sigma^{\mu}_{a\bar{b}} \partial_{\mu} \bar{\xi}^{(j)\bar{b}}
= m \xi^{(j)}_{a} \qquad
- i~\sigma^{\mu}_{a\bar{b}} \partial_{\mu} \xi^{(j)a}
= m \bar{\xi}^{(j)}_{\bar{b}}
\eea
and the usual causal anti-commutation relations:
\bea
\left\{ \xi^{(j)}_{a}(x), \xi^{(k)}_{b}(y) \right\}
= i~\delta_{jk}~\epsilon_{ab}~m~D_{m}(x-y)
\nonumber \\
\left\{ \xi^{(j)}_{a}(x), \bar{\xi}^{(k)}_{\bar{b}}(y) \right\} =
\delta_{jk}~\sigma^{\mu}_{a\bar{b}}~\partial_{\mu}D_{m}(x-y)
\eea
and conversely, if two spinors
$\xi^{(j)}_{a}, \quad j = 1,2$
verify the preceding relations, then the spinor
\be
f_{a} \equiv {1\over 2m} (\xi^{(2)}_{a} - i~\xi^{(1)}_{a})
\ee
verifies Klein-Gordon equation and the causal anticommutation relation
(\ref{CCR-d}). The correspondence
$f_{a} \leftrightarrow (\xi^{(j)}_{a})_{j=1}^{2}$
is one-one.
\label{dirac-majorana}
\end{prop}

The proof is elementary and shows that every (four-component) Dirac bi-spinor
can be described by a Weyl spinor verifying only Klein-Gordon equation; such
a spinor can in turn be described by two Majorana spinors. It follows that
we do not lose generality if we consider all Fermi fields of spin $1/2$ to
be Majorana spinors.

Now we define the notion of supersymmetry invariance of the system of Bosonic
and Fermionic fields considered above. Suppose that in the Hilbert space
${\cal H}$
we also have the operators
$Q_{a}, \quad a = 1,2$
such that:

(i) the following relations are verified:
\be
Q_{a} \Omega = 0, \quad \bar{Q}_{\bar{a}} \Omega = 0
\label{vac}
\ee
and
\bea
\{ Q_{a} , Q_{b} \} = 0, \quad
\{ Q_{a} , \bar{Q}_{\bar{b}} \} = 2 \sigma^{\mu}_{a\bar{b}} P_{\mu}, \quad
[ Q_{a}, P_{\mu} ] = 0, \quad
U_{A}^{-1} Q_{a} U_{A} = {A_{a}}^{b} Q_{b}.
\label{SUSY}
\eea

Here
$P_{\mu}$
are the infinitesimal generators of the translation group given by
\be
[P_{\mu}, b ] = - i~\partial_{\mu} b, \quad
[P_{\mu}, f ] = - i~\partial_{\mu} f.
\label{P}
\ee
and
\be
\bar{Q}_{\bar{b}} \equiv (Q_{b})^{\dagger}.
\ee

(ii) The following commutation relations are true:
\bea
i [ Q_{a}, b ] = p(\partial) f,
\qquad
\{ Q_{a}, f \} = q(\partial) b
\label{tensor}
\eea
where
$b = (b^{(j)}, b_{\mu})$
(resp.
$f = (f^{(A)}_{a}, \bar{f}^{(A)}_{\bar{a}}$)
is the collection of all integer (resp. half-integer) spin fields and $p, q$
are matrix-valued polynomials in the partial derivatives
$\partial_{\mu}$
(with constant
 coefficients). These relations express the tensor properties
of the fields 
with respect to (infinitesimal) supersymmetry transformations.

If this conditions are true we say that
$Q_{a}$
are {\it super-charges} and
$b, f$
are forming a {\it supersymmetric multiplet}. The notion of
{\it irreducibility} can be defined for any supersymmetric multiplet if
we consider the quantum fields as a modulus over the ring of partial
differential operators. As emphasised in \cite{GS1}, the matrix-valued
operators
 $p$ and $q$
are subject to various constraints. Let us describe them in this context.
\begin{itemize}
\item
>From the compatibility of
(\ref{tensor}) with Lorentz transformations it
follows that these polynomials
are Lorentz covariant.
\item
Next, we start from the fact that the Hilbert space of the model is
generated by vectors of the type
\bea
\Psi = \prod b(x_{p})~\prod f(x_{q})~\Omega \in {\cal H}.
\label{hilbert}
\eea
The action of the supercharges
$Q_{a}, \quad \bar{Q}_{\bar{a}}$
is determined by (\ref{tensor}): one commutes the supercharge operators to
the right till they hit the vacuum and then one applies (\ref{vac}).
However, the supercharges are not independent: they are constrained by the
relations from (\ref{SUSY}) and we should check that we do not get
a contradiction. The consistency relations are given by the (graded) Jacobi
identities combined with (\ref{SUSY}) and the relation (\ref{P}):

As a result we must have:
\bea
\left\{ Q_{a} , [ Q_{b}, b ] \right\} = - ( a \leftrightarrow b)
\nonumber \\
\left[ Q_{a} , \{ Q_{b}, f \} \right] = - ( a \leftrightarrow b), \qquad
\nonumber \\
\left\{ Q_{a} , [ \bar{Q}_{\bar{b}}, b ] \right\}
+ \{ \bar{Q}_{\bar{b}} , [ Q_{a}, b ] \} =
- 2i~\sigma^{\mu}_{a\bar{b}}~\partial_{\mu} b,
\nonumber \\
\left[ Q_{a} , \{ \bar{Q}_{\bar{b}}, f \} \right]
+ [ \bar{Q}_{\bar{b}} , \{ Q_{a}, f \} ] =
- 2i~\sigma^{\mu}_{a\bar{b}}~\partial_{\mu} f.
\label{susy+lorentz}
\eea
\item
The equation of motion (\ref{kg-1}), (\ref{kg-2}) and (\ref{dirac}) are
supersymmetric invariant, i.e. if we take the commutator of the supercharges
$Q_{a}$
and
$\bar{Q}_{\bar{a}}$
with the equations (\ref{kg-1}) and (\ref{kg-2}) we obtain zero modulo
(\ref{dirac}); also if we take the anticommutator of the equations of motion
(\ref{dirac}) with the supercharges
$Q_{a}, \bar{Q}_{\bar{a}}$
we get zero modulo (\ref{kg-1}), (\ref{kg-2}).
\item
The (anti)commutation relations have the implication that one and the same
vector from the Hilbert space
${\cal H}$
can be expressed in the form (\ref{hilbert}) in two distinct ways. This means
that the supercharges are well defined {\it via} (\ref{SUSY}) {\it iff} some
new consistency relations are valid following again from graded Jacobi
identities; the non-trivial ones are of the form:
\bea
[ b(x), \{ f(y), Q_{a} \} ] = - \{ f(y), [ {Q}_{a}, b(x) ] \}
\label{susy+CCR}
\eea
\item
If a gauge supercharge $Q$ is present in the model, then it is usually
determined by relations of the type (\ref{tensor}) involving ghost fields also,
so it means that we must impose consistency relations of the same type as
above. Moreover, it is desirable to have
\be
\{ Q, Q_{a} \} = 0, \quad \{ Q, \bar{Q}_{\bar{a}} \} = 0;
\label{susy+gauge}
\ee
this implies that the supersymmetric charges
$Q_{a}$
and
$\bar{Q}_{\bar{a}}$
factorizes to the physical Hilbert space
${\cal H}_{phys} = Ker(Q)/Im(Q)$.
This implies new consistency relations of the type (\ref{susy+lorentz})
with one of the supercharges replaced by the gauge charge:
\bea
~\left\{ Q_{a} , [ Q, b ] \right\} = - \left\{ Q , [ Q_{a}, b ] \right\}
\qquad
~\left[ Q_{a} , \{ Q, f \} \right] = - \left[ Q, \{ Q_{a}, f \} \right].
\label{susy+gauge1}
\eea
\item
A relation of the type (\ref{susy+CCR}) must be also valid for the
gauge charge:
\bea
[ b(x), \{ f(y), Q \} ] = - \{ f(y), [ {Q}, b(x) ] \}
.
\label{susy+CCR-Q}
\eea
\item
To have
$Q^{2} = 0$
we must also impose
\be
~\{ Q, [Q , b ] \} = 0 \qquad [ Q, \{ Q, f \} ] = 0.
\label{q2}
\ee
\end{itemize}
\begin{rem}
All these conditions are of pure quantum nature i.e. they can
 be understood
only for a pure quantum model. Some of them do not have a classical analogue
so we can interpret the obstacles in constructing supersymmetric quantum
models (associated to some classical supersymmetric theories) as some
quantum anomalies.
\end{rem}

It seems to be an essential point to describe supersymmetric theories in
{\it superspace} \cite{SS1}, \cite{SS2}. We do this in the following way.
We consider the space
${\cal H}_{G} \equiv {\cal G} \otimes {\cal H}$
where
${\cal G}$
is a Grassmann algebra generated by Weyl anticommuting spinors
$\theta_{a}$
and their complex conjugates
$\bar{\theta}_{\bar{a}} = (\theta_{a})^{*}$
and perform a Klein transform such that the Grassmann parameters
$\theta_{a}$
are anti-commuting with all Fermionic fields, the supercharges and
the gauge charge. The field operators acting in
${\cal H}_{G}$
are called {\it superfields}. Of special interest are the superfields
constructed as in \cite{CS}, \cite{CGS} according to the formul\ae:
\bea
B(x,\theta,\bar{\theta}) \equiv
W_{\theta,\bar{\theta}}~b(x)~W_{\theta,\bar{\theta}}^{-1},
\nonumber \\
F(x,\theta,\bar{\theta}) \equiv
W_{\theta,\bar{\theta}}~f(x)~W_{\theta,\bar{\theta}}^{-1},
\label{superfields}
\eea
where
\be
W_{\theta,\bar{\theta}} \equiv
\exp\left(i\theta^{a} Q_{a} - i\bar{\theta}^{\bar{a}} \bar{Q}_{\bar{a}}\right)
\label{W-expo}
\ee
and we interpret the exponential as a (finite) Taylor series.

It is a 
remarkable fact that only such type of superfields are really
necessary,
 so in the following, when referring to superfields we mean
expressions given
 by (\ref{superfields}). We will call them {\it super-Bose}
and respectively 
 {\it super-Fermi} fields.
 For convenience we will denote
frequently the ensemble of 
Minkowski and Grassmann variables by
$X = (x,\theta,\bar{\theta})$.

More generally one starts from Wick monomials defined in
${\cal H}$
and by multiplication with Grassmann variables we obtain
super-Wick monomials in the extended Fock space
${\cal H}_{G}$.
It appeared from the analysis of \cite{GS1} that it is worthwhile to define
a canonical map associating to any Wick monomial $w$ in
${\cal H}$
a {\it super-Wick monomials} acting in
${\cal H}_{G}$
according to the formula:
\bea
(sw)(x,\theta,\bar{\theta}) \equiv
W_{\theta,\bar{\theta}}~w(x)~W_{\theta,\bar{\theta}}^{-1}
;
\label{s}
\eea
(here $s$ stands for ``sandwich formula" or for ``supersymmetric extension".)
From now on by supersymmetric Wick monomials we mean only expressions of the
type
$sw$.

Now we have some elementary results from \cite{GS1} which will be repeatedly
used in the computations; for simplicity we  denote by $[ , ]$ the graded
commutator.
\begin{lemma}
(i) Let
 us define the operators:
\bea
D_{a} \equiv {\partial \over \partial \theta^{a}}
+ i \sigma^{\mu}_{a\bar{b}} \bar{\theta}^{\bar{b}} \partial_{\mu}
\qquad
\bar{D}_{\bar{a}} \equiv - {\partial \over \partial \bar{\theta}^{\bar{a}}}
- i \sigma^{\mu}_{b\bar{a}} \theta^{b} \partial_{\mu}.
\label{D}
\eea

Then if
$W \equiv s(w)$
the following formul\ae~ are true:
\bea
i [ Q_{a}, W(x,\theta,\bar{\theta}) ] = D_{a} W(x,\theta,\bar{\theta})
\qquad
i [ \bar{Q}_{\bar{a}}, W(x,\theta,\bar{\theta}) ] =
\bar{D}_{\bar{a}} W(x,\theta,\bar{\theta})
\label{BCH-inf}
\eea

(ii) Let us define the operators
\bea
{\cal D}_{a} \equiv {\partial \over \partial \theta^{a}}
- i \sigma^{\mu}_{a\bar{b}} \bar{\theta}^{\bar{b}} \partial_{\mu}
\qquad
\bar{\cal D}_{\bar{a}} \equiv
- {\partial \over \partial \bar{\theta}^{\bar{a}}}
+ i \sigma^{\mu}_{b\bar{a}} \theta^{b} \partial_{\mu}
\label{calD}
\eea
acting on any superfield (or super--Wick polynomials). Then for any Wick
monomial
$w(x)$
the following relations are true:
\be
{\cal D}_{a} sw = i~s ([Q_{a}, w]), \qquad
\bar{\cal D}_{\bar{a}} sw = i~s([\bar{Q}_{\bar{a}}, w]).
\label{d+s}
\ee

(iii) The operators
$D_{a}$
and
$\bar{D}_{\bar{a}}$
verify the following formul\ae:
\bea
( D_{a} T)^{\dagger} = \pm \bar{D}_{\bar{a}} T^{\dagger},
\nonumber \\
\{D_{a}, D_{b} \} = 0, \quad
\{\bar{D}_{\bar{a}}, \bar{D}_{\bar{b}} \} = 0, \quad
\{D_{a}, \bar{D}_{\bar{b}} \} = -2 i \sigma^{\mu}_{a\bar{b}}~\partial_{\mu}
\label{DD}
\eea
where in the first formula the sign $+ (-)$ corresponds to a super-Bose (-Fermi)
field. The operators
${\cal D}$
verify relations of the same type.
\end{lemma}

Let us comment on the physical interpretation of the forml\ae~ (\ref{BCH-inf}
).
If the Wick polynomial
$W(x,\theta,\bar{\theta})$
verifies (\ref{BCH-inf}
) then let us define
\be
w(x) \equiv \int d\theta^{2} d\bar{\theta}^{2} W(x,\theta,\bar{\theta});
\label{WW}
\ee
it follows from (\ref{BCH-inf}
) that we have
\bea
~[ Q_{a}, w(x) ] = i \partial_{\mu}w_{a}^{\mu}(x)
\qquad
~[ \bar{Q}_{\bar{a}}, w(x) ] = 
 i \partial_{\mu}w_{a}^{\mu}(x)
^{\dagger}
\label{BCH-x}
\eea
where
\be
w_{a}^{\mu} \equiv \sigma^{\mu}_{a\bar{b}} \int d\theta^{2} d\bar{\theta}^{2}
\bar{\theta}^{\bar{b}}~W(x,\theta,\bar{\theta});
\ee
the equations (\ref{BCH-x}) are exactly the supersymmetry postulate used
in \cite{Gr2}. It is not clear if the converse is true i.e. suppose we have
(\ref{BCH-x}) for some Wick polynomial $w$; then is it possible to find out
a supersymmetric Wick polynomial $W$ such that we have (\ref{WW}) and
(\ref{BCH-inf})?

Concerning the meaning of (\ref{d+s}) let us consider for an arbitrary
supersymmetric Wick monomial
$W(x,\theta,\bar{\theta})$
the operation of restriction to the ``initial value" (in the Grassmann
variables):
\be
(rW)(x) \equiv W(x,0,0).
\label{r}
\ee
Then the formul\ae~ (\ref{d+s}) imply
\be
{\cal D}_{a} W = i~s ([Q_{a}, rW]), \qquad
\bar{\cal D}_{\bar{a}} W = i~s([\bar{Q}_{\bar{a}}, rW]).
\label{s+r}
\ee

These equations can be regarded as a system of  partial
differential equations (in the Grassmann variables) and this system
determines uniquely the supersymmetric Wick monomials $W$ if one knows
the ``initial values"
$w = rW$.
(If there are two solutions, then their difference verifies the
associated homogeneous equation which tells that there is no dependence on
the Grassmann variables; but the ``initial values" for the difference is zero.)

For another point of view concerning supersymmetric Hilbert spaces we refer
to the recent paper \cite{Ru}.

We close this Section mentioning that for the construction of the ghost and
anti-ghost multiplets (which are needed in the construction of supersymmetric
gauge
 theories) one must consider that the integer (resp. half-integer) spin
fields have Fermi-Dirac (resp. Bose-Einstein) statistics and to invert
everywhere the r\^ole of commutators and anti-commutators. The scheme
presented above is reproduced with minimal changes.

\section{The Vector Multiplet\label{vector}}

By definition, the {\it vector} multiplet has the content described in
Section \ref{qsr}:  the Bosonic fields are some (real) scalars
$b^{(j)}, \quad j = 1,\dots,s$
and a real vector field
$b_{\mu}$;
the Fermionic fields are some Majorana fields of spin one-half
$f^{(A)}, \quad A = 1,\dots,f$.
Let us consider
\be
C \equiv \sum_{j} \gamma_{j}~b^{(j)}
\label{C}
\ee
for some real constants
$\gamma_{j}$,
not all of them zero (i.e. in vector notations
$\vec{\gamma} \not= \vec{0}$
because otherwise we would have
$C = 0$
also). In particular $C$ can be one of the scalar fields
$b^{(j)}$.
Now we define the following superfield:
$
V \equiv s(C)
$
i.e.
\be
V(x,\theta,\bar{\theta}) \equiv
W_{\theta,\bar{\theta}}~C(x)~W_{\theta,\bar{\theta}}^{-1}.
\ee

It is clear that one has the reality condition
\be
V^{\dagger} = V;
\ee
this is not a restriction: is $V$ is not Hermitian we consider its Hermitian
and anti-Hermitian parts separately.

Moreover the generic expression of $V$ must be
\bea
V(x,\theta,\bar{\theta}) = C(x) + \theta\chi(x) + \bar{\theta} \bar{\chi}(x)
+ \theta^{2}~\phi(x) + \bar{\theta}^{2}~ \phi^{\dagger}
(x)
\nonumber \\
+ (\theta \sigma^{\mu} \bar{\theta})~v_{\mu}(x)
+ \theta^{2}~\bar{\theta}\bar{\lambda}(x)
+ \bar{\theta}^{2}~\theta\lambda(x)
+ \theta^{2} \bar{\theta}^{2}~ d(x)
\label{V}
\eea
where, from Lorentz covariance arguments, we must have:
\bea
\chi = \sum_{A} \alpha_{A}~f^{(A)} \qquad
\lambda = \sum_{A} \beta_{A}~f^{(A)}
\nonumber \\
d = \sum_{j} \delta_{j}~b^{(j)} + \delta_{0}~\partial^{\mu}b_{\mu}
\nonumber \\
\phi = \sum_{j} \rho_{j}~b^{(j)} + \rho_{0}~\partial^{\mu}b_{\mu}
\nonumber \\
v_{\mu} = \tau_{0}~b_{\mu} + \sum_{j} \tau_{j}~\partial_{\mu}b^{(j)}
\label{comp}
\eea
for some (complex) numbers
$
\vec{\alpha}, \vec{\beta}, \vec{\delta}, \vec{\rho}, \vec{\tau},
\delta_{0}, \rho_{0}, \tau_{0}.
$
>From the reality condition we must have:
\be
d^{\dagger} = d, \qquad v_{\mu}^{\dagger} = v_{\mu}
\ee
so
$
\vec{\delta}, \vec{\tau}, \delta_{0}, \tau_{0}
$
must be real.

Now we determine the action of the supercharges on the components of the
multiplet.
\begin{prop}
In the preceding conditions, the following relations are true:
\bea
i~[Q_{a}, C ] = \chi_{a}
\nonumber \\
~\{ Q_{a}, \chi_{b} \} = 2i~\epsilon_{ab} \phi
\nonumber \\
\{ Q_{a}, \bar{\chi}_{\bar{b}} \} = - i~\sigma^{\mu}_{a\bar{b}}
~
( v_{\mu} + i~\partial_{\mu}C )
\nonumber \\
~[Q_{a}, \phi ] = 0
\nonumber \\
i~[Q_{a}, \phi^{\dagger} ] = \lambda_{a} - {i\over 2} \sigma^{\mu}_{a\bar{b}}
\partial_{\mu}\bar{\chi}^{\bar{b}}
\nonumber \\
i~[Q_{a}, v^{\mu} ] = \sigma^{\mu}_{a\bar{b}} \bar{\lambda}^{\bar{b}}
- {i\over 2} \partial^{\mu}\chi_{a}
- \sigma^{\mu\rho}_{ab} \partial_{\rho}\chi^{b}
\nonumber \\
~\{ Q_{a}, \lambda_{b} \} =
i~\epsilon_{ab} \left( 2 d + {i\over 2} \partial^{\mu}v_{\mu} \right)
- i~\sigma^{\mu\rho}_{ab} \partial_{\mu}v_{\rho}
\nonumber \\
\{ Q_{a}, \bar{\lambda}_{\bar{b}} \}
= \sigma^{\mu}_{a\bar{b}}
\partial_{\mu}\phi
\nonumber \\
~[Q_{a}, d ] =
- {1\over 2} \sigma^{\mu}_{a\bar{b}}\partial_{\mu}\bar{\lambda}^{\bar{b}}
\label{susy-action}
\eea

These relations are compatible with the Jacoby identities (\ref{susy+lorentz}
).

If $C$ verifies the Klein-Gordon equation (for mass $m$)
\be
(\partial^{2} + m^{2}) C = 0
\ee
then the superfield $V$ verifies the Klein-Gordon equation
\be
(\partial^{2} + m^{2}) V = 0
\ee
so all the components of the multiplet are verifying Klein-Gordon equation
of mass $m$. These equations are compatible with the supersymmetry action i.e.
they are left invariant by the supercharges
$Q_{a}$
and
$\bar{Q}_{\bar{a}}$.

The multiplet
$(C, \phi, v_{\mu}, d, \lambda_{a}, \chi_{a})$
is irreducible; in particular it follows that the indices $j$ and $A$ take
four values, $C$ and $d$ are real scalar fields of mass $m$,
$b_{\mu}$
is a real vector field of mass $m$,
$\phi$
is a complex scalar field of mass $m$ and
$\chi_{a}$
and
$\lambda_{a}$
are Dirac fields of mass $m$ (both of them being equivalent to a pair of
Majorana fields in the sense of proposition \ref{dirac-majorana}).
\end{prop}

{\bf Proof:}
We use the relation (\ref{BCH-inf})
\bea
i [ Q_{a}, V(x,\theta,\bar{\theta}) ] = D_{a} V(x,\theta,\bar{\theta})
\qquad
i [ \bar{Q}_{\bar{a}}, V(x,\theta,\bar{\theta}) ] =
\bar{D}_{\bar{a}} V(x,\theta,\bar{\theta})
\label{bch}
\eea
and if we introduce in both hand sides the expression (\ref{V}) of the vector
superfield we obtain by straightforward computations the action of the
supercharges (\ref{susy-action}). The verification of the relations
(\ref{susy+lorentz}) is long but also straightforward. One can avoid this
long computations if one derives immediately from the preceding relation that:
\bea
~\{  Q_{a} , [ Q_{b}, V ] \} = - ( a \leftrightarrow b)
\nonumber \\
~\{  Q_{a} , [ \bar{Q}_{\bar{b}}, V ] \} -
\{ \bar{Q}_{\bar{b}} , [ Q_{a}, V ] \}  =
- 2i~\sigma^{\mu}_{a\bar{b}}~\partial_{\mu} V
;
\eea
if we introduce here the expression (\ref{V}) of  $V$ and consider the various
coefficients of the Grassmann variables, then one obtains (\ref{susy+lorentz}).

The assertion concerning the Klein-Gordon equation is immediate. The
irreducibility of the multiplet follows by {\it reductio ad absurdum}. One
admits that a relation of the type
\be
\alpha~C + \beta~\phi + \bar{\beta}~\phi^{\dagger}
+ \gamma~\partial^{\mu}v_{\mu} + \delta~d = 0
\ee
is true (this being the most general linear dependence between the Bosonic
fields compatible with Lorentz covariance; higher derivatives are excluded
if one uses Klein-Gordon and Dirac equations). Then one commutes twice with the
supercharges and discovers some contradictions. If one has a relations between
the Fermi fields, the the anticommutator with the supercharges gives a relation
of the preceding type.
$\qed$
\begin{rem}
The relations (\ref{susy-action}) are, essentially, those from the literature
- see for instance \cite{GGRS} formula (3.6.5) - in our notations. If one
makes the change of fields \cite{Sr}
\bea
\lambda^{\prime}_{a} \equiv \lambda_{a} + {i\over 2} \sigma^{\mu}_{a\bar{b}}
\partial_{\mu}\bar{\chi}^{\bar{b}}
\nonumber \\
d^{\prime} \equiv d - {m^{2} \over 4} C
\eea
then the (\ref{susy-action}) acquires a somewhat simple form:
\bea
i~[Q_{a}, C ] = \chi_{a}
\nonumber \\
~\{ Q_{a}, \chi_{b} \} = 2i~\epsilon_{ab} \phi
\nonumber \\
\{ Q_{a}, \bar{\chi}_{\bar{b}} \} = - i~\sigma^{\mu}_{a\bar{b}}
~
( v_{\mu} + i~\partial_{\mu}C )
\nonumber \\
~[Q_{a}, \phi ] = 0
\nonumber \\
i~[Q_{a}, \phi^{\dagger} ] = \lambda^{\prime}_{a}
- i~\sigma^{\mu}_{a\bar{b}}\partial_{\mu}\bar{\chi}^{\bar{b}}
\nonumber \\
i~[Q_{a}, v^{\mu} ] = \sigma^{\mu}_{a\bar{b}} \bar{\lambda^{\prime}}^{\bar{b}}
- i~\partial^{\mu}\chi_{a}
\nonumber \\
~\{ Q_{a}, \lambda^{\prime}_{b} \} = 2i~\epsilon_{ab} d^{\prime}
- 2i~\sigma^{\mu\rho}_{ab} \partial_{\mu}v_{\rho}
\nonumber \\
\{ Q_{a}, \bar{\lambda}^{\prime}_{\bar{b}} \} = 0
\nonumber \\
~[Q_{a}, d^{\prime} ] =
- {1\over 2} \sigma^{\mu}_{a\bar{b}}
\partial_{\mu}\bar{\lambda^{\prime}}^{\bar{b}}.
\eea
\end{rem}

It is clear that the one-particle Hilbert space of the vector multiplet is
bigger than the representation
$\Omega_{1/2}$
described by formula (\ref{decomp}). To obtain  the ``physical" Fock space
associated to
$\Omega_{1/2}$
we have to follow the idea outlined in the Introduction, namely to extend
the Hilbert space of the vector field with some superghost and anti-superghost
fields and find a gauge charge operator $Q$ such that the ``physical" Fock space
is given by the formula
$Ker(Q)/ Im(Q)$.
A hint about this construction is given by
\begin{prop}
The vector superfield $V$ can be written as follows
\be
V = \sum_{j=0}^{2} P_{j} V
\ee
where the expressions
$P_{j}, \quad j = 0,1,2$
are given by
\be
P_{0} \equiv - {1\over 8m^{2}} {\cal D}^{a} \bar{\cal D}^{2} {\cal D}_{a},
\quad
P_{1} \equiv {1\over 16m^{2}} {\cal D}^{2} \bar{\cal D}^{2}, \quad
P_{2} \equiv {1\over 16m^{2}} \bar{\cal D}^{2} {\cal D}^{2}.
\ee

The expressions
$P_{j}, \quad j = 0,\dots,2$
are projectors on the mass shell i.e. they verify
\bea
P_{j} P_{k} = 0, \quad \forall j \not= k,
\nonumber \\
P_{j}^{2}~V = P_{j}~V, \quad \forall j.
\eea

The components
$V_{j} \equiv P_{j}~V, \quad j =1,2$
of $V$ verify
\be
{\cal D}_{a}~V_{1} = 0, \quad \bar{\cal D}_{a}~V_{2} = 0.
\label{chiral-dec}
\ee
\label{decomp-v}
\end{prop}

{\bf Proof:}
The proof follows from some elementary identities verified by
${\cal D}_{a}$
and
$\bar{\cal D}_{\bar{a}}$:
\bea
{\cal D}^{2} \bar{\cal D}^{2} + \bar{\cal D}^{2} {\cal D}^{2}
- 2~{\cal D}^{a} \bar{\cal D}^{2} {\cal D}_{a} = - 16 \partial^{2},
\nonumber \\
({\cal D}^{2} \bar{\cal D}^{2})^{2}
= - 16 \partial^{2}~{\cal D}^{2} \bar{\cal D}^{2}, \qquad
(\bar{\cal D}^{2} {\cal D}^{2})^{2}
= - 16 \partial^{2}~\bar{\cal D}^{2} {\cal D}^{2}
\nonumber \\
{\cal D}_{a} {\cal D}_{b} {\cal D}_{c} = 0, \qquad
\bar{\cal D}_{\bar{a}} \bar{\cal D}_{\bar{b}} \bar{\cal D}_{\bar{c}} = 0.
\eea

The proof of these identities is elementary (see for instance \cite{Sr}).
$\qed$

The relations (\ref{chiral-dec}) are called {\it chirality} (resp.
{\it anti-chirality}) conditions.

We now have directly from (\ref{d+s}) the following
\begin{prop}
Let us define
\bea
V_{\mu} \equiv s(v_{\mu}) \qquad
D \equiv s(d) \qquad D^{\prime} \equiv s(d^{\prime}) \qquad \Phi \equiv s(\phi)
\nonumber \\
X_{a} \equiv s(\chi_{a}), \qquad \Lambda_{a} \equiv s(\lambda_{a}) \qquad
\Lambda^{\prime}_{a} \equiv s(\lambda^{\prime}_{a}).
\label{asoc}
\eea

Then the following relations are true:
\bea
{\cal D}_{a} V = X_{a}
\nonumber \\
{\cal D}_{a} X_{b} = - 2 \epsilon_{ab} \Phi
\nonumber \\
{\cal D}_{a} \bar{X}_{\bar{b}}  = \sigma^{\mu}_{a\bar{b}}
~
( V_{\mu} + i~\partial_{\mu}V )
\nonumber \\
{\cal D}_{a} \Phi  = 0
\nonumber \\
{\cal D}_{a} \Phi^{\dagger}  = \Lambda^{\prime}_{a} - i~\sigma^{\mu}_{a\bar{b}}
\partial_{\mu}\bar{X}^{\bar{b}}
\nonumber \\
{\cal D}_{a} V^{\mu}  = \sigma^{\mu}_{a\bar{b}} \bar{\Lambda^{\prime}}^{\bar{b}}
- i~\partial^{\mu}X_{a}
\nonumber \\
{\cal D}_{a} \Lambda^{\prime}_{b} = - 2 \epsilon_{ab} D^{\prime}
+ 2 \sigma^{\mu\rho}_{ab} \partial_{\mu}V_{\rho}
\nonumber \\
{\cal D}_{a} \bar{\Lambda}^{\prime}_{\bar{b}} = 0
\nonumber \\
{\cal D}_{a} D^{\prime}  =
- {i\over 2} \sigma^{\mu}_{a\bar{b}}
\partial_{\mu}\bar{\Lambda^{\prime}}^{\bar{b}}.
\label{dv}
\eea
\end{prop}

Using these relations one can express all associated superfields from
(\ref{asoc}) as some polynomial in the operators
${\cal D}_{a}$
and
$\bar{\cal D}_{\bar{a}}$
applied to $V$; in particular the following algebraic relations:
\bea
\Phi = - {1\over 4} {\cal D}^{2}V
\quad
\partial_{\mu}V^{\mu} = {i\over 16} [ \bar{\cal D}^{2}, {\cal D}^{2} ]V
\quad
D = {1\over 32} \left({\cal D}^{2} \bar{\cal D}^{2}
+ \bar{\cal D}^{2} {\cal D}^{2} \right) V - {m^{2} \over 4}~V
\label{algebra}
\eea
can be obtained. One can determine by direct computation the expression of
the superfield
$V_{0}$:
\bea
V_{0} = - {2\over m^{2}}~D^{\prime}
\nonumber \\
D^{\prime} = d^{\prime} -
{i\over 2}~\theta \sigma^{\mu} \partial_{\mu}\bar{\lambda}^{\prime}
+ {i\over 2}~\partial_{\mu}\lambda^{\prime} \sigma^{\mu} \bar{\theta}
- {1\over 2}~(\theta \sigma^{\mu} \bar{\theta})~( m^{2}~g_{\mu\rho}
+ \partial_{\mu}\partial_{\rho} )v^{\rho}
\nonumber \\
- {m^{2} \over 4}~\left(\theta^{2}~\bar{\theta}\bar{\lambda}^{\prime}
+ \bar{\theta}^{2}~\theta\lambda^{\prime} \right)
- {m^{2} \over 4}~\theta^{2} \bar{\theta}^{2}~d'
\label{d-prime}
\eea
so the superfield
$V_{0}$
contains only one Majorana spinor
$\lambda^{\prime}$,
one scalar field
$d^{\prime}$
and a real vector field
\be
v^{\prime}_{\mu} \equiv
\left(g_{\mu\rho} + {1\over m^{2}} \partial_{\mu}\partial_{\rho}\right)v^{\rho}
\ee
verifying the transversality condition
\be
\partial^{\mu}v^{\prime}_{\mu} = 0.
\ee

If one applies the superfield
$V_{0}(x)$
to the vacuum one can see that
$\Omega_{1/2}$
is generated. So, to obtain the ``physical" Hilbert space associated to
$\Omega_{1/2}$
one has to eliminate the chiral and the anti-chiral parts of $V$. As above,
one can determine by direct computation that the superfields
$V_{1}$
and
$V_{2}$
do not contain particles of spin $1$; indeed the field
$v_{\mu}$
appears in the two superfields only through the combination
$\partial^{\mu}v_{\mu}$.

We now determine the supercommutator of the vector field. We have the following
result.
\begin{thm}
The vector multiplet exists only for
$m > 0$.
In this case, the generic form of the causal (anti)commutators of the fields
are:
\bea
~[ C(x), C(y) ] = - i~D_{m}(x-y)
\nonumber \\
~[ C(x), d(y) ] = - i~\alpha~D_{m}(x-y)
\nonumber \\
~[ C(x), \phi(y) ] = - i~\beta~D_{m}(x-y)
\nonumber \\
~[ \phi(x), \phi^{\dagger}(y) ] =
- i~\left(\alpha + {m^{2}\over 4}\right)~D_{m}(x-y)
\nonumber \\
~[ \phi(x), d(y) ] = {m^{2}\beta\over 4}~D_{m}(x-y)
\nonumber \\
~[ \phi(x), v_{\mu}(y) ] = i~\beta~\partial_{\mu}D_{m}(x-y)
\nonumber \\
~[ d(x), d(y) ] = - {im^{4}\over 16}~D_{m}(x-y)
\nonumber \\
~[ v_{\mu}(x), v_{\rho}(y) ] =
i~\partial_{\mu}\partial_{\rho}~D_{m}(x-y)
+ i~\left({m^{2}\over 2} - 2\alpha \right)~g_{\mu\rho}~D_{m}(x-y)
\nonumber \\
\left\{ \chi_{a}(x), \chi_{b}(y) \right\}
= 2 \beta~\epsilon_{ab}~D_{m}(x-y),
\nonumber \\
\left\{ \chi_{a}(x), \bar{\chi}_{\bar{b}}(y) \right\} =
\sigma^{\mu}_{a\bar{b}}~\partial_{\mu}D_{m}(x-y)
\nonumber \\
\left\{ \lambda_{a}(x), \lambda_{b}(y) \right\}
= - {m^{2}\bar{\beta}\over 2}~\epsilon_{ab}~D_{m}(x-y),
\nonumber \\
\left\{ \lambda_{a}(x), \bar{\lambda}_{\bar{b}}(y) \right\} =
{m^{2}\over 4}~\sigma^{\mu}_{a\bar{b}}~\partial_{\mu}D_{m}(x-y)
\nonumber \\
\left\{ \chi_{a}(x), \lambda_{b}(y) \right\}
= - 2i \alpha~\epsilon_{ab}~D_{m}(x-y),
\nonumber \\
\left\{ \chi_{a}(x), \bar{\lambda}_{\bar{b}}(y) \right\} =
i \beta~\sigma^{\mu}_{a\bar{b}}~\partial_{\mu}D_{m}(x-y)
\label{CCR-V}
\eea
and all other (anti)commutators are zero. Here
$\alpha \in \R$
and
$\beta \in \C$
are two free parameters constrained only by the inequalities
\be
|\alpha| \leq {m^{2} \over 4} \quad |\beta| \leq {m \over 2} \quad
|Im(\beta)| \leq {m \over 4} + {\alpha \over 4}.
\ee
\end{thm}

{\bf Proof:}
Starting directly from (\ref{CCR-1}) + (\ref{CCR-2}) + (\ref{CCR-3}) and
(\ref{C}) + \ref{comp}) we immediately get:
\bea
~[ C(x), C(y) ] = - i~|\vec{\gamma}|^{2}~D_{m}(x-y)
\nonumber \\
~[ C(x), d(y) ] = - i~\vec{\gamma}\cdot\vec{\delta}~D_{m}(x-y)
\nonumber \\
~[ C(x), \phi(y) ] = - i~\vec{\gamma}\cdot\vec{\rho}~D_{m}(x-y)
\nonumber \\
~[ C(x), v_{\mu}(y) ] =- i~\vec{\gamma}\cdot\vec{\tau}~\partial_{\mu}D_{m}(x-y)
\nonumber \\
~[ \phi(x), \phi(y) ] =
- i~\left(|\vec{\rho}|^{2} - m^{2} \rho_{0}^{2}\right)~D_{m}(x-y)
\nonumber \\
~[ \phi(x), \phi^{\dagger}(y) ] = - i~\left(\vec{\rho}\cdot\vec{\rho}^{*}
- m^{2} |\rho_{0}|^{2}\right)~D_{m}(x-y)
\nonumber \\
~[ \phi(x), d(y) ] = - i~\left(\vec{\delta}\cdot\vec{\rho}
- m^{2} \delta_{0}\rho_{0}\right)~D_{m}(x-y)
\nonumber \\
~[ \phi(x), v_{\mu}(y) ] = i~\left(\vec{\rho}\cdot\vec{\tau}
- m^{2} \rho_{0} \tau_{0}\right)~D_{m}(x-y)
\nonumber \\
~[ d(x), d(y) ] =
- i~\left(|\vec{\delta}|^{2} - m^{2} \delta_{0}^{2}\right)~D_{m}(x-y)
\nonumber \\
~[ d(x), v_{\mu}(y) ] =
i~\left(\vec{\delta}\cdot\vec{\tau} - \delta_{0} \tau_{0}\right)~D_{m}(x-y)
\nonumber \\
~[ v_{\mu}(x), v_{\rho}(y) ] = i~\left(\tau_{0}^{2}~g_{\mu\rho}
+ |\vec{\tau}|^{2}\partial_{\mu}\partial_{\rho}\right)~D_{m}(x-y)
\nonumber \\
\left\{ \chi_{a}(x), \chi_{b}(y) \right\}
= i~m~\vec{\alpha}^{2}~\epsilon_{ab}~D_{m}(x-y),
\nonumber \\
\left\{ \chi_{a}(x), \bar{\chi}_{\bar{b}}(y) \right\} =
\vec{\alpha}\cdot\vec{\alpha}^{*}
\sigma^{\mu}_{a\bar{b}}~\partial_{\mu}D_{m}(x-y)
\nonumber \\
\left\{ \lambda_{a}(x), \lambda_{b}(y) \right\}
= i~m~|\vec{\beta}|^{2}~\epsilon_{ab}~D_{m}(x-y),
\nonumber \\
\left\{ \lambda_{a}(x), \bar{\lambda}_{\bar{b}}(y) \right\} =
\vec{\beta}\cdot\vec{\beta}^{*}
\sigma^{\mu}_{a\bar{b}}~\partial_{\mu}D_{m}(x-y)
\nonumber \\
\left\{ \chi_{a}(x), \lambda_{b}(y) \right\}
= i~m~\vec{\beta}\cdot\vec{\alpha}~\epsilon_{ab}~D_{m}(x-y),
\nonumber \\
\left\{ \chi_{a}(x), \bar{\lambda}_{\bar{b}}(y) \right\} =
\vec{\alpha}\cdot\vec{\beta}^{*}
\sigma^{\mu}_{a\bar{b}}~\partial_{\mu}D_{m}(x-y)
\label{CCR-V-generic}
\eea

Now we consider the first equation. Because
$\vec{\gamma} \not= \vec{0}$
we can rescale $C$ (and implicitly $V$) and make
$|\vec{\gamma}| = 1$;
in this way we arrange that we have the first equation of (\ref{CCR-V}).
Next we define the real numbers
$\alpha, \gamma$
and the complex number
$\beta$
according to
\be
\alpha \equiv \vec{\gamma}\cdot\vec{\delta} \qquad
\beta \equiv \vec{\gamma}\cdot\vec{\rho} \qquad
\gamma \equiv \vec{\gamma}\cdot\vec{\tau}.
\ee

If we consider all non-trivial Jacobi identities (\ref{susy+CCR}) we obtain
after some computation that
$\gamma = 0$
and the rest of the relations of (\ref{CCR-V}).

If
$m = 0$
we get from (\ref{CCR-V})
$
\left\{ \lambda_{a}(x), \bar{\lambda}_{\bar{b}}(y) \right\} = 0
$
so using (\ref{CCR-V-generic}) we get
$
\vec{\beta}\cdot\vec{\beta}^{*} = 0;
$
this implies
$\vec{\beta} = \vec{0}$
so
$\lambda_{a} = 0$,
absurd. So we must have
$m > 0$.
In this case the inequalities from the statement follow from the
Cauchy-Schwartz inequalities: one splits all complex vectors in the real and
complex part and  considers all pair of vectors so obtained.
$\qed$

We now give an alternative way of computing the causal (anti)commutation
relations having the advantage of being more abstract and giving directly
the causal super-commutator of the vector superfield. We first have:
\begin{prop}
Let us consider the causal commutator
\be
~[ V(X_{1}), V(X_{2}) ] = - i D(X_{1};X_{2})~ {\bf 1}.
\ee

Then the expression
$D(X_{1};X_{2})$
is a distribution in the variables
$x_{j}$
and a polynomial in the Grassmann variables
$\theta_{j}, \quad j = 1,2$
and verifies the following properties:

(a) it is Poincar\'e covariant; in particular it depends only on the difference
$x_{1} - x_{2}$;

(b) it has causal support;

(c) verifies Klein-Gordon equation:
\be
(\partial^{2} + m^{2} ) D(X_{1};X_{2}) = 0;
\ee

(d) verifies the Hermiticity condition:
\be
\overline{D(X_{1};X_{2})} = - D(X_{2};X_{1});
\ee

(e) verifies the antisymmetry condition:
\be
D(X_{2};X_{1}) = - D(X_{1};X_{2}).
\ee

(f) verifies the consistency condition:
\be
(D^{1}_{a} + D^{2}_{a} ) D(X_{1};X_{2}) = 0;
\ee
\label{super-causal}
\end{prop}

{\bf Proof:}
All properties except (f) are immediate. If we start from the Jacobi identity
\be
~[ Q_{a}, [ V(X_{1}), V(X_{2}) ] ] + [  V(X_{1}), [ V(X_{2}),  Q_{a} ]]
+ [ V(X_{2}), [ Q_{a},  V(X_{1}) ] ]  = 0
\ee
and we use (\ref{bch}) to obtain
\be
(D^{1}_{a} + D^{2}_{a} )~[ V(X_{1}), V(X_{2}) ]  = 0
\ee
and (f) follows.
$\qed$

\begin{prop}
The generic solution of the conditions (a)-(f) of the preceding proposition is
\be
D(X_{1};X_{2}) = \exp [i\left( \theta_{1} \sigma^{\mu} \bar{\theta}_{2}
- \theta_{2} \sigma^{\mu} \bar{\theta}_{1} \right) \partial_{\mu} ]~
E(\theta_{1} - \theta_{2}; \bar{\theta}_{1} - \bar{\theta}_{2}; x_{1} - x_{2})
\ee
where the expression $E$ is a distribution in the variable
$x$
and a polynomial in the Grassmann variables
$\zeta \equiv \theta_{1} - \theta_{2}$
and verifies the following properties:

(a) it is Lorentz covariant;

(b) it has causal support;

(c) verifies Klein-Gordon equation:
\be
(\partial^{2} + m^{2} ) E = 0;
\ee

(d) verifies the Hermiticity condition:
\be
\overline{E(\zeta;\bar{\zeta};x)} = - E(-\zeta;-\bar{\zeta};-x);
\ee

(e) verifies the antisymmetry condition:
\be
E(\zeta;\bar{\zeta};x) = - E(-\zeta;-\bar{\zeta};-x);
\ee
\end{prop}

{\bf Proof:}
One rewrites the consistency condition (f) in the variables
\be
\theta \equiv {1\over 2} (\theta_{1} + \theta_{2} ), \quad
\zeta = \theta_{1} - \theta_{2}
\ee
and obtains
\be
\left( {\partial \over \partial \theta^{a}}
+ i~\sigma^{\mu}_{a\bar{b}} \bar{\zeta}^{\bar{b}} \partial_{\mu} \right) D = 0.
\label{f}
\ee

This equation can be ``integrated" to
\be
D(\theta,\bar{\theta};\zeta;\bar{\zeta};x) =
\exp [ i\left( \zeta\sigma^{\mu} \bar{\theta}
- \theta\sigma^{\mu} \bar{\zeta} \right) \partial_{\mu} ]~
E(\zeta;\bar{\zeta};x);
\ee
to be completely rigorous one has to expand $D$ as a polynomial in
$\theta$
and
$\bar{\theta}$
and to translate the equation (\ref{f}) in relations between the coefficients.
Now it is easy to express the properties (a) to (e) of the preceding
proposition in terms of $E$ and to revert to the old variables.
$\qed$
\begin{prop}
The solutions of the problem from proposition \ref{super-causal} generate a
real vector space of dimension 4; a basis in this space can be taken to be:
\bea
D_{1}(X_{1};X_{2}) = \exp [i\left( \theta_{1} \sigma^{\mu} \bar{\theta}_{2}
- \theta_{2} \sigma^{\mu} \bar{\theta}_{1} \right) \partial_{\mu} ]~
D_{m}(x_{1}-x_{2})
\nonumber \\
D_{2}(X_{1};X_{2}) =
(\theta_{1}-\theta_{2})^{2} (\bar{\theta}_{1}-\bar{\theta}_{2})^{2}~
D_{m}(x_{1}-x_{2})
\nonumber \\
D_{3}(X_{1};X_{2}) = \exp [i\left( \theta_{1} \sigma^{\mu} \bar{\theta}_{2}
- \theta_{2} \sigma^{\mu} \bar{\theta}_{1} \right) \partial_{\mu} ]
[ (\theta_{1}-\theta_{2})^{2}  + (\bar{\theta}_{1}-\bar{\theta}_{2})^{2} ]~
D_{m}(x_{1}-x_{2})
\nonumber \\
D_{4}(X_{1};X_{2}) = i~\exp [i\left( \theta_{1} \sigma^{\mu} \bar{\theta}_{2}
- \theta_{2} \sigma^{\mu} \bar{\theta}_{1} \right) \partial_{\mu} ]
[ (\theta_{1}-\theta_{2})^{2} - (\bar{\theta}_{1}-\bar{\theta}_{2})^{2} ]~
D_{m}(x_{1}-x_{2})
\label{basis}
\eea

In particular the expressions
$D_{j}$
do not contain terms with odd number of Grassmann variables. In consequence
any Bose field are commuting with any Fermi field.
\end{prop}

{\bf Proof:}
The generic form of $E$ is, from Lorentz covariance considerations:
\be
E = A_{1} + \zeta^{2} A_{2} - \bar{\zeta}^{2} A_{3}
+ \zeta\sigma^{\mu}\bar{\zeta} \partial_{\mu}A_{4}
+ \zeta^{2} \bar{\zeta}^{2} A_{5}
\ee
where
$A_{j}, \quad j = 1,\dots,5$
are numerical distributions. One can impose now the restrictions (a) to (e) from
the preceding proposition and gets quite easily the result from the statement.
$\qed$

It follows that the general solution of the problem (a) to (f) from the
proposition \ref{super-causal}) is of the form:
\be
D(X_{1};X_{2}) = \sum_{j=1}^{4} c_{j}~D_{j}(X_{1};X_{2}).
\label{ddd}
\ee

After some computations one can match this expression with (\ref{CCR-V}); we
have:
\be
c_{1} = 1, \quad \alpha = c_{2}, \quad \beta = c_{3} - i~c_{4}
\label{abc}
\ee
so the super-order of singularity is:
\be
\omega(D(X_{1};X_{2})) = -2.
\ee

We also give some interesting relations verified by the expressions
$D_{j}(X_{1};X_{2})$.
We have:
\begin{prop}
The following relations are true:
\bea
{\cal D}^{2}~D_{1} = - {m^{2} \over 2} ( D_{3} + i~D_{4} )
\nonumber \\
{\cal D}^{2}~D_{2} = - 2 ( D_{3} + i~D_{4} )
\nonumber \\
{\cal D}^{2}~D_{3} = - 4~D_{1} - m^{2}~D_{2} - 4i~D_{5}
\nonumber \\
{\cal D}^{2}~D_{4} = - 4i~ D_{1} - i~m^{2}~D_{2} + 4~D_{5}
\nonumber \\
{\cal D}^{2}~D_{5} = - i~m^{2} ( D_{3} + i~~D_{4} )
\nonumber \\
\eea
where we have defined
\be
D_{5}(X_{1};X_{2}) = \exp [i\left( \theta_{1} \sigma^{\mu} \bar{\theta}_{2}
- \theta_{2} \sigma^{\mu} \bar{\theta}_{1} \right) \partial_{\mu} ]~
(\theta_{1}-\theta_{2}) \sigma^{\rho} (\bar{\theta}_{1}-\bar{\theta}_{2})
\partial_{\rho}D_{m}(x_{1}-x_{2})
\ee
and the operator
${\cal D}^{2}$
pertains to the variable
$X_{1}$.
\label{dd}
\end{prop}

{\bf Proof:}
If we consider only the conditions (a) to (d) from the proposition
\ref{super-causal} we obtain a complex vector space of dimension 5 with a basis
given by
$D_{j}, \quad j = 1,\dots,5$.
Now it is clear that the expressions
${\cal D}^{2}~D_{j}$
are also verifying the conditions (a) to (d), so we must have relations of the
following type:
\be
{\cal D}^{2}~D_{j} = \sum_{k=1}^{5} c_{jk}~D_{k}
\ee
for some complex coefficients
$c_{jk}$.
To determine these coefficients we make
$\theta_{2} \rightarrow 0, \bar{\theta}_{2} \rightarrow 0$
and some simple computations.
$\qed$

In perturbation theory we need the expression of the Feynman super-propagator.
This can be obtained from the expression of the super-causal distribution $D$
by distribution splitting \cite{Sc}, \cite{Gr1}; in this simple case this
amounts to make the
replacement
\be
D_{m}(x) \quad \longrightarrow \quad D_{m}^{F}(x)
\label{feynman}
\ee
where at the right-hand side we have the usual expression of the Feynman
propagator. This means that we have
\be
D^{F}(X_{1};X_{2}) = \sum_{j=1}^{4} c_{j}~D^{F}_{j}(X_{1};X_{2}).
\ee
where the expressions
$D^{F}_{j}(X_{1};X_{2})$
are obtained from (\ref{basis}) with the substitution (\ref{feynman}).

Let us emphasise now an important departure from the standard literature. The
expression of the super-propagator appearing in the standard literature is
$D^{F}_{2}(X_{1};X_{2})$
- see for instance \cite{FP} formula (5.23). But one can immediately see that
the choice
$c_{1} = 0$
is in conflict with the basic theorem about the structure of the super-causal
distribution $D$. Indeed, if the causal super-commutator is
$D_{2}$
this means in particular that we have
$
[ C(x), C(y) ] = 0
$
which implies
$
C = 0, \quad V = 0
$.
So, it seems that the formal manipulations based on the formal path integral
integration are not completely safe: one can obtain in this way formal
theories which do not have a {\it bona~ fid\ae} representation in a Hilbert
space. These theories are in obvious conflict with good old fashion quantum
mechanics!
 Another departure from the standard literature is the proof
that the model exists only for
$m > 0$.

\section{Chiral Multiplets\label{chiral-multi}}

\subsection{The Chiral Scalar Multiplet\label{scalar}}

The decomposition given by Proposition \ref{decomp-v} shows that the
ghost and the antighost superfields should be constrained by two restrictions:
they should not generate spin $1$ particles and they should obey the chirality
condition. These two conditions determines what it is called in the
literature the scalar chiral superfield. (We will also need the ghost version of
such a superfield.) By definition a {\it scalar chiral superfield} is a
superfield
$H(x,\theta,\bar{\theta})$
verifying the following conditions: (a) it corresponds to a multiplet of fields
of spin
$s \leq 1/2$;
(b) it is a scalar with respect to Poincar\'e transformations; (c) it verifies
the {\it chirality}  condition:
\be
{\cal D}_{a}~H = 0;
\ee
the {\it anti-chirality} condition is obviously
\be
\bar{\cal D}_{\bar{a}}~H = 0.
\ee
If
$H$
is a chiral superfield, then
$H^{\dagger}$
is an anti-chiral superfield and vive-versa so we can study only one of them,
say chiral superfields. It is easy to determine the generic form of a chiral
superfield; from Lorentz covariance we get
\bea
H(x,\theta,\bar{\theta}) = h(x)
+ 2~\bar{\theta} \bar{\psi}(x)
+ i~(\theta \sigma^{\mu} \bar{\theta})~\partial_{\mu}h(x)
+ \bar{\theta}^{2}~f(x)
- i~\bar{\theta}^{2}~\theta \sigma^{\mu} \partial_{\mu}\bar{\psi}(x)
+ {m^{2} \over 4}~\theta^{2} \bar{\theta}^{2}~h(x)
\label{chiral}
\eea
where
$h, f$
and
$\psi$
have expressions of the type (\ref{comp}):
\bea
h = \sum_{j} \rho_{j}~b^{(j)} \qquad
f = \sum_{j} \delta_{j}~b^{(j)} \qquad
\psi = \sum_{A} \alpha_{A}~f^{(A)}
\label{comp-chiral}
\eea
for some real scalar fields
$b^{(j)}$
and some Majorana fields
$f^{(A)}$.
These are free fields i.e. Klein-Gordon, respectively Dirac equations are
verified. We proceed as for the vector multiplet.
\begin{prop}
Let us suppose that
$H = s(h)$.
Then the action of the supercharges on the chiral multiplet,
consistent with the conditions (\ref{susy+lorentz}) is:
\bea
i~[Q_{a}, h ] = 0 \qquad
i~[Q_{a}, h^{\dagger} ] = 2\psi_{a}
\nonumber \\
~[Q_{a}, f ] = -2~\sigma^{\mu}_{a\bar{b}}
~\partial_{\mu}\bar{\psi}^{\bar{b}}
\qquad
~[Q_{a}, f^{\dagger} ] = 0
\nonumber \\
~\{ Q_{a}, \psi_{b} \} = i~\epsilon_{ab} f^{\dagger}
\qquad
\{ Q_{a}, \bar{\psi}_{\bar{b}} \} = \sigma^{\mu}_{a\bar{b}}
~\partial_{\mu}h.
\label{susy-ch}
\eea
\label{susy-chiral}

As a result, if the field
 $h$
 is of mass $m$ all the fields of the multiplet
must be of mass $m$.
\end{prop}
{\bf Proof:}
As for the vector superfield we have from (\ref{BCH-inf})
\bea
i [ Q_{a}, H(x,\theta,\bar{\theta}) ] = D_{a} H(x,\theta,\bar{\theta})
\qquad
i [ \bar{Q}_{\bar{a}}, H(x,\theta,\bar{\theta}) ] =
\bar{D}_{\bar{a}} H(x,\theta,\bar{\theta})
\eea
and if we introduce the expression (\ref{chiral}) we get the action of the
supercharges. The consistency conditions (\ref{susy+lorentz}) are verified by
direct computation. The last assertion is elementary.
$\qed$

If we do not impose additional constraints, then we must have
 in the
general scheme from Section \ref{qsr} (namely the relations
(\ref{h-1}) - (\ref{CCR-3}))
$s = 4$
and
$f = 2$
so the chiral multiplet is reducible: we have twice the representation
\be
\Omega_{0} \sim [m,0] \oplus [m,0] \oplus [m,1/2]
\ee
of the super-Poincar\'e group. Also from (\ref{d+s})
 we have
\begin{prop}
Let us define
\be
F \equiv s(f), \qquad \Psi_{a} \equiv s(\psi_{a}).
\ee

Then the following relations are true:
\bea
{\cal D}_{a} H = 0
\qquad
{\cal D}_{a} H^{\dagger}  = 2~\Psi_{a}
\nonumber \\
{\cal D}_{a} F = -2i~\sigma^{\mu}_{a\bar{b}}
~\partial_{\mu}\bar{\Psi}^{\bar{b}}
\qquad
{\cal D}_{a} F^{\dagger}  = 0
\nonumber \\
{\cal D}_{a} \Psi_{b} = -~\epsilon_{ab} F^{\dagger}
\qquad
{\cal D}_{a} \bar{\Psi}_{\bar{b}}
= i~\sigma^{\mu}_{a\bar{b}}
~\partial_{\mu}H.
\eea

As a consequence we have
\bea
\bar{\cal D}^{2} H = - 4~F \qquad
{\cal D}^{2} F = - 4m^{2}~H \qquad
{\cal D}^{2} \Psi_{a} = 0.
\eea

In particular we have the ``equation of motion"
\be
{\cal D}^{2} \bar{\cal D}^{2} H = 16m^{2}~H
.
\ee
\end{prop}

It was proved in \cite{LW} (see also \cite{GS1}) that any supersymmetric
multiplet of fields with spin
$s \leq 1/2$
is a sum of {\it Wess-Zumino} multiplets \cite{WZ}. By definition, 
a
Wess-Zumino multiplet is composed from a complex scalar field
$h$
and a spin $1/2$ Majorana field
$f_{a}$
of the same mass $m$. In particular
$f_{a}$
verifies Dirac equation. The relations (\ref{tensor}) are in this case by
definition:
\bea
~[Q_{a}, h ] = 0
\qquad
i~[Q_{a}, h^{\dagger} ] = 2 f_{a}
\nonumber \\
~\{ Q_{a}, f_{b} \} = - i~m~\epsilon_{ab} h, \qquad
~\{ Q_{a}, \bar{f}_{\bar{b}} \} = \sigma^{\mu}_{a\bar{b}} \partial_{\mu}h
\label{wess}
\eea

This multiplet is irreducible. The causal (anti)commutators are:
\bea
\left[ h(x), h(y)^{\dagger} \right] = - 2i\alpha~ D_{m}(x-y),
\nonumber \\
\left\{ f_{a}(x), f_{b}(y) \right\} = i~~\alpha~\epsilon_{ab}~m~D_{m}(x-y),
\nonumber \\
\left\{ f_{a}(x), \bar{f}_{\bar{b}}(y) \right\}
= \alpha~\sigma^{\mu}_{a\bar{b}}~\partial_{\mu}D_{m}(x-y)
\label{CCR-wess}
\eea
and the other (anti)commutators are zero; here
$\alpha \in \R^{+}$
is a arbitrary parameter. One can prove them if one starts from the first
relation and uses the consistency conditions (\ref{susy+CCR}).

We now show that the chiral multiplet is a sum of two Wess-Zumino multiplets.
\begin{prop}
In the conditions of proposition 
\ref{susy-chiral} let us define
\bea
\psi^{(+)}_{a} \equiv
i~\sigma^{\mu}_{a\bar{b}} \partial_{\mu}\bar{\psi
}^{\bar{b}} + m\psi_{a} \qquad
h^{(+)} \equiv m h - f^{\dagger}
\nonumber \\
\psi^{(-)}_{a} \equiv
\sigma^{\mu}_{a\bar{b}} \partial_{\mu}\bar{\psi
}^{\bar{b}} + i~m\psi_{a} \qquad
h^{(-)} \equiv i~(m h + f^{\dagger}).
\eea

The the couples
$(h^{(\pm)}_{a},\psi^{(\pm)})$
are two Wess-Zumino multiplets.
\label{chiral=2wz}
\end{prop}
{\bf Proof:}
By direct computation one proves that for both couples the relations
(\ref{wess}) are verified. Then one notices that
\be
h = {1\over 2m} \left[ h^{(+)} + i~h^{(-)} \right] \qquad
f = {1\over 2} \left[ h^{(-)} - h^{(+)} \right] \qquad
\psi_{a} = {1\over 2m} \left[ \psi^{(+)}_{a} - i~\psi^{(-)}_{a} \right]
\ee
so the correspondence
$
(\psi_{a},h)  \leftrightarrow (\psi^{(\pm)}_{a},h^{(\pm)})
$
is one-one.
$\qed$

One usually obtains the Wess-Zumino multiplet from the chiral multiplet by
imposing
\be
\psi^{(-)}_{a} = 0 \qquad h^{(-)} = 0
\ee
which are equivalent to the {\it equation of motion} namely:
\bea
\bar{\cal D}^{2} H = 4m H^{\dagger}.
\label{super-eq}
\eea

The determination of the causal (anti)commutation relation for the chiral
multiplet follows the usual pattern.
\begin{prop}
The generic causal (anti)commutators for the chiral multiplet are
\bea
~[ h(x), h^{\dagger}(y) ] = - i\alpha^{\prime}~D_{m}(x-y)
\nonumber \\
~[ h(x), f(y) ] = \beta^{\prime}~D_{m}(x-y)
\nonumber \\
~[ f(x), f^{\dagger}(y) ] = - i~m^{2}~\alpha^{\prime}~D_{m}(x-y)
\nonumber \\
\left\{ \psi_{a}(x), \psi_{b}(y) \right\}
= - {\bar{\beta}^{\prime} \over 2} ~\epsilon_{ab}~D_{m}(x-y),
\nonumber \\
\left\{ \psi_{a}(x), \bar{\psi}_{\bar{b}}(y) \right\} = {\alpha^{\prime}\over 2}
\sigma^{\mu}_{a\bar{b}}~\partial_{\mu}D_{m}(x-y)
\label{CCR-chiral}
\eea
and the other (anti)commutators are zero. Here
$\alpha^{\prime} \in \R^{+}, \quad \beta^{\prime} \in \C$
are arbitrary parameters.
\end{prop}
{\bf Proof:}
We start with the first two relations which can be justified from
(\ref{comp-chiral}). If we use the consistency conditions (\ref{susy+CCR})
we obtain the other (anti)commutators.
$\qed$
\begin{rem}
If require that the two Wess-Zumino multiplets associated to the chiral
multiplet in the sense of the proposition \ref{chiral=2wz} are completely
decoupled i.e. they (anti)commute one with the other, then we must impose
$Re(\beta) = 0$
.
However, the general consistency conditions from Section \ref{qsr} are valid
for arbitrary
$\beta$
so we will not restrict this parameter.
\end{rem}

There is an alternative expression of the chiral superfield (\ref{chiral})
\bea
H(x,\theta,\bar{\theta})
= \exp (\theta\sigma^{\mu}\bar{\theta} \partial_{\mu})~
[ h(x) + 2~\bar{\theta} \bar{\psi}(x) + \bar{\theta}^{2}~f(x)
 ]
\label{chiral2}
\eea
which can be used to determine the commutation relations of the superfields:
\bea
\left[ H(X_{1}),
 H(X_{2}) \right] 
= \beta^{\prime}~D_{+}(X_{1};X_{2})
\nonumber \\
~\left[ H(X_{1}), H^{\dagger}(X_{2}) \right]
= -~i~\alpha^{\prime}~D_{-}(X_{1};X_{2})
\label{commutators}
\eea
where
\bea
D_{+}(X_{1};X_{2}) = (\bar{\theta}_{1} - \bar{\theta}_{2})^{2}
\exp[ i~(\theta_{1} \sigma^{\mu} \bar{\theta}_{2}
- \theta_{2} \sigma^{\mu} \bar{\theta}_{1}) \partial_{\mu} ]~
D_{m}(x_{1} - x_{2})
\nonumber \\
D_{-}(X_{1};X_{2}) = \exp[ i~(\theta_{1} \sigma^{\mu} \bar{\theta}_{1}
+ \theta_{2} \sigma^{\mu} \bar{\theta}_{2}
- 2~\theta_{2} \sigma^{\mu} \bar{\theta}_{1} ) \partial_{\mu} ]~
D_{m}(x_{1} - x_{2})
\label{d-pm}
\eea
or, if we use the causal super-distribution
$D_{j},~j = 1,\dots,5$
introduced in the preceding Section:
\bea
D_{+} = {1\over 2}  ( D_{3} -i~D_{4} )
\qquad
D_{-} = D_{1} - {m^{2}\over 2} D_{2} + i~D_{5}.
\eea

The super-order of singularities are better than the general formula
 from
\cite{GS1}, namely:
\bea
\omega(D_{-}) = - 2, \qquad \omega(D_{+}) = - 3.
\eea

We end this subsection with a critical analysis of the so-called Wess-Zumino
gauge. It is asserted in the literature (see for instance \cite{Li}
) that
one can write a vector superfield in the form
\be
V = V_{WZ} + H + H^{\dagger}
\label{dec}
\ee
where $H$ is a chiral superfield and
$V_{WZ}$
has the generic form
\bea
V_{WZ}(x,\theta,\bar{\theta}) =
(\theta \sigma^{\mu} \bar{\theta})~w_{\mu}(x)
+ \theta^{2}~\bar{\theta}\bar{\omega}(x)
+ \bar{\theta}^{2}~\theta\omega(x)
+ \theta^{2} \bar{\theta}^{2}~v(x)
;
\label{V-wz}
\eea
here $v$ is a real scalar fields,
$w_{\mu}$
is a real vector field and
$\omega_{a}$
is a Dirac spinor.
If one introduces in (\ref{dec}) the explicit expressions (\ref{V}) and
(\ref{chiral}) then one immediately get the relations
\be
C = h + h^{\dagger} \qquad \chi_{a} = 2 \psi_{a} \qquad \phi = f^{\dagger}.
\ee

We show that the decomposition (\ref{dec}) is {\bf not} supersymmetric
invariant. For this we simply take the (anti)commutators of the supercharges
$Q_{a}$
and
$\bar{Q}_{\bar{a}}$
with the three preceding relations. In particular we must have
\be
~[ \bar{Q}_{\bar{a}}, \phi - f^{\dagger} ] = 0.
\ee

But if we use (\ref{susy-action}) and (\ref{susy-ch}) 
we immediately get
\be
\lambda^{\prime}_{a} = 0
\ee
i.e. a contradiction. The conclusion is that the Wess-Zumino gauge is not
a legitimate supersymmetric decomposition: if one supposes that $V$ is a
superfields then $H$ cannot be a superfield and vice-versa.

\subsection{Chiral Ghost and Antighost Multiplets}

To define in
 a consistent super-symmetric way ghost and anti-ghost fields
one only has to invert the statistics assignment: we assume that the ghost
multiplet is build from some scalar
 fields
$u^{(j)}$
which are Hermitian and respect Fermi-Dirac statistics; their Majorana
partners
$f^{(A)}_{a}$
are Bosons. The anti-ghost multiplet has a similar structure but we change
the Hermiticity properties in agreement to the usual conventions \cite{Sc}:
the scalar
 fields
$\tilde{u}^{(j)}$
are anti-Hermitian and respect Fermi-Dirac statistics; their 
anti-Majorana
partners
$\tilde{f}^{(A)}_{a}$
are Bosons:
\bea
(u^{(j)})^{\dagger} = u^{(j)} \qquad
(f^{(A)}_{a})^{\dagger} = \bar{f}^{(A)}_{\bar{a}}
\nonumber \\
(\tilde{u}^{(j)})^{\dagger} =  - \tilde{u}^{(j)}
\qquad
(\tilde{f}^{(A)}_{a})^{\dagger} = - \bar{\tilde{f}}^{(A)}_{\bar{a}}.
\eea

These are free fields i.e. Klein-Gordon, respectively Dirac equations are
verified. Now the changes in the preceding arguments are minimal.
It is easy to determine the generic form of a chiral ghost and anti-ghost
superfields; we get
\bea
U(x,\theta,\bar{\theta}) = u(x)
+ 2i~\bar{\theta} \bar{\zeta}(x)
+ i~(\theta \sigma^{\mu} \bar{\theta})~\partial_{\mu}u
+ \bar{\theta}^{2}~g(x)
+ \bar{\theta}^{2}~\theta \sigma^{\mu} \partial_{\mu}\bar{\zeta}(x)
+ {m^{2} \over 4}~\theta^{2} \bar{\theta}^{2}~u(x)
\label{U}
\eea
and respectively
\bea
\tilde{U}(x,\theta,\bar{\theta}) = \tilde{u}(x)
- 2i~\bar{\theta} \bar{\tilde{\zeta}}(x)
+ i~(\theta \sigma^{\mu} \bar{\theta})~\partial_{\mu}\tilde{u}
+ \bar{\theta}^{2}~\tilde{g}(x)
- \bar{\theta}^{2}~\theta \sigma^{\mu} \partial_{\mu}\bar{\tilde{\zeta}}(x)
+ {m^{2} \over 4}~\theta^{2} \bar{\theta}^{2}~\tilde{u}(x)
\label{tildeU}
\eea
where
$u, g$
(resp.
$\tilde{u}, \tilde{g}$)
are linear combinations of
$u^{(j)}$
(resp. of
$\tilde{u}^{(j)}$ )
and
$\zeta$
(resp.
$\tilde{\zeta}$ )
are linear combinations of
$f^{(A)}$
(resp.
$\tilde{f}^{(A)}$).

Instead of the formul\ae~of the proposition \ref{susy-chiral} we get:
\bea
\{ Q_{a}, u \} = 0 \qquad
\{ Q_{a}, u^{\dagger} \} = 2\zeta_{a}
\nonumber \\
\{ Q_{a}, g \} = -2i~\sigma^{\mu}_{a\bar{b}}
~\partial_{\mu}\bar{\zeta}^{\bar{b}}
\qquad
\{ Q_{a}, g^{\dagger} \} = 0
\nonumber \\
~[ Q_{a}, \zeta_{b} ] = \epsilon_{ab} g^{\dagger}
\qquad
i [ Q_{a}, \bar{\zeta}_{\bar{b}} ] = \sigma^{\mu}_{a\bar{b}}
~\partial_{\mu}u
\eea
and respectively:
\bea
\{ Q_{a}, \tilde{u} \} = 0 \qquad
\{ Q_{a}, \tilde{u}^{\dagger} \} = 2\tilde{\zeta}_{a}
\nonumber \\
\{ Q_{a}, \tilde{g} \} = 2i~\sigma^{\mu}_{a\bar{b}}
~
\partial_{\mu}\bar{\tilde{\zeta}}^{\bar{b}}
\qquad
\{ Q_{a}, \tilde{g}^{\dagger} \} = 0
\nonumber \\
~[ Q_{a}, \tilde{\zeta}_{b} ] = \epsilon_{ab} \tilde{g}^{\dagger}
\qquad
i [ Q_{a}, \bar{\tilde{\zeta}}_{\bar{b}} ] = - \sigma^{\mu}_{a\bar{b}}
~
\partial_{\mu}\tilde{u}.
\eea
In particular all fields of the same multiplet must have the same mass.

If we define
\be
G \equiv s(g), \qquad Z_{a} \equiv s(\zeta_{a})
\label{gz}
\ee
then we have:
\bea
{\cal D}_{a} U = 0
\qquad
{\cal D}_{a} U^{\dagger}  = 2i~Z_{a}
\nonumber \\
{\cal D}_{a} G = 2~\sigma^{\mu}_{a\bar{b}}
~\partial_{\mu}\bar{Z}^{\bar{b}}
\qquad
{\cal D}_{a} G^{\dagger}  = 0
\nonumber \\
{\cal D}_{a} Z_{b} = i~\epsilon_{ab} G^{\dagger}
\qquad
{\cal D}_{a} \bar{Z}_{\bar{b}}
= \sigma^{\mu}_{a\bar{b}}
~\partial_{\mu}U.
\eea
and as a consequence we have
\bea
\bar{\cal D}^{2} U = - 4~G \qquad
{\cal D}^{2} G = - 4m^{2}~U \qquad
{\cal D}^{2} Z_{a} = 0.
\eea
In particular we have the ``equation of motion"
\be
{\cal D}^{2} \bar{\cal D}^{2} U = 16m^{2}~U
.
\ee

Analogously if we define
\be
\tilde{G} \equiv s(\tilde{g}), \qquad \tilde{Z}_{a} \equiv s(\tilde{\zeta}_{a})
\label{gz-tilde}
\ee
then we have:
\bea
{\cal D}_{a} \tilde{U} = 0
\qquad
{\cal D}_{a} \tilde{U}^{\dagger}  = 2i~\tilde{Z}_{a}
\nonumber \\
{\cal D}_{a} \tilde{G} =
- 2~\sigma^{\mu}_{a\bar{b}}
~\partial_{\mu}\bar{\tilde{Z}}^{\bar{b}}
\qquad
{\cal D}_{a} \tilde{G}^{\dagger}  = 0
\nonumber \\
{\cal D}_{a} \tilde{Z}_{b} = i~\epsilon_{ab} \tilde{G}^{\dagger}
\qquad
{\cal D}_{a} \bar{\tilde{Z}}_{\bar{b}}
= - \sigma^{\mu}_{a\bar{b}}
~\partial_{\mu}\tilde{U}.
\eea
and as a consequence we have
\bea
\bar{\cal D}^{2} \tilde{U} = - 4~\tilde{G} \qquad
{\cal D}^{2} \tilde{G} = - 4m^{2}~\tilde{U} \qquad
{\cal D}^{2} \tilde{Z}_{a} = 0.
\eea
In particular we have the ``equation of motion"
\be
{\cal D}^{2} \bar{\cal D}^{2} \tilde{U} = 16m^{2}~\tilde{U}
.
\ee

The ghost multiplet is a sum of elementary
 ghost multiplets built from a
complex scalar field $u$ with Fermi statistics and a Majorana spinor
$\psi$
with Bose statistics of the same mass
$m$
such that we have instead of (\ref{wess}) the following
 relations \cite{GS1}:
\bea
~\{ Q_{a}, u \} = 0, \qquad
\{ Q_{a}, u^{\dagger} \} = 2 \zeta_{a}
\nonumber \\
~[ Q_{a}, \zeta_{b} ] = - m~\epsilon_{ab} u \qquad
i~[ Q_{a}, \bar{\zeta}_{\bar{b}} ] = \sigma^{\mu}_{a\bar{b}} \partial_{\mu}u
\label{wess-gh}
\eea
and for the anti-ghost multiplet of mass $m$ we have instead:
\bea
~\{ Q_{a}, \tilde{u} \} = 0 \qquad
~\{ Q_{a}, \tilde{u}^{\dagger} \} = 2 \tilde{\zeta}_{a}
\nonumber \\
~[ Q_{a}, \tilde{\zeta}_{b} ] = m~\epsilon_{ab} \tilde{u}, \qquad
i~[ Q_{a}, \bar{\tilde{\zeta}}_{\bar{b}} ]
= - \sigma^{\mu}_{a\bar{b}} \partial_{\mu}\tilde{u}
\label{wess-antigh}
\eea

The decomposition of the chiral ghost and antighost multiplets into irreducible
ones goes as above and the formul\ae~are:
\bea
\zeta^{(+)}_{a} \equiv
i~\sigma^{\mu}_{a\bar{b}} \partial_{\mu}\bar{\zeta
}^{\bar{b}} + m\zeta_{a}
\qquad
u^{(+)} \equiv m u - g^{\dagger}
\nonumber \\
\zeta^{(-)}_{a} \equiv
\sigma^{\mu}_{a\bar{b}} \partial_{\mu}\bar{\zeta
}^{\bar{b}} + i~m\zeta_{a}
\qquad
u^{(-)} \equiv - i~(m u + g^{\dagger})
\eea
and respectively:
\bea
\tilde{\zeta}^{(+)}_{a} \equiv
i~\sigma^{\mu}_{a\bar{b}} \partial_{\mu}\bar{\tilde{\psi}
}^{\bar{b}}
+ m\tilde{\zeta}_{a} \qquad
\tilde{u}^{(+)} \equiv - (m \tilde{u} + \tilde{g}^{\dagger} )
\nonumber \\
\tilde{\zeta}^{(-)}_{a} \equiv
\sigma^{\mu}_{a\bar{b}} \partial_{\mu}\bar{\tilde{\psi}
}^{\bar{b}}
+ i~m\tilde{\zeta}_{a} \qquad
u^{(-)} \equiv i~(m \tilde{u} - \tilde{g}^{\dagger}).
\eea

The reduction of the chiral ghost and antighost multiplets can be done imposing
supersymmetric equations of motion:
\bea
\bar{\cal D}^{2} U = 4m~U^{\dagger}
\qquad
\bar{\cal D}^{2} \tilde{U} = 4m~\tilde{U}^{\dagger}.
\label{super-eq-gh}
\eea

Up to now, the ghost and the antighost multiplets can be considered of distinct
masses. However to consider the commutation relations we remember that for
usual gauge theories \cite{Sc} one has to consider that the ghost and the
anti-ghost fields are of the same mass and verify commutation relations
of
 the following type:
\bea
\left\{ u^{(j)}(x), \tilde{u}^{(k)}(y) \right\} = -~i~\delta_{jk}~D_{m}(x-y)
\nonumber \\
~[ f^{(A)}_{a}(x), \tilde{f}^{(B)}_{b}(y) ]
= -~i~\delta_{AB}~\epsilon_{ab}~D_{m}(x-y)
\nonumber \\
~[ f^{(A)}_{a}(x), \bar{\tilde{f}}^{(B)}_{\bar{b}}(y) ] = -~\delta_{AB}~
\sigma^{\mu}_{a\bar{b}}\partial_{\mu}D_{m}(x-y).
\eea

Then we have from (\ref{susy+CCR-Q}) the generic causal (anti)commutator
relations for the chiral multiplets:
\bea
~\{ u(x), \tilde{u}^{\dagger}(y) \} = i\alpha^{\prime\prime}~D_{m}(x-y)
\nonumber \\
~\{ u(x), \tilde{g}(y) \} = \beta^{\prime\prime}~D_{m}(x-y)
\nonumber \\
~\{ g(x), \tilde{g}^{\dagger}(y) \} = i~m^{2}~\alpha^{\prime\prime}~D_{m}(x-y)
\nonumber \\
~\{ g(x), \tilde{u}(y) \} = \beta^{\prime\prime}~D_{m}(x-y)
\nonumber \\
~[ \zeta_{a}(x), \tilde{\zeta}_{b}(y) ] =
{\bar{\beta}^{\prime\prime} \over 2}~\epsilon_{ab}~D_{m}(x-y)
\nonumber \\
~[ \zeta_{a}(x), \bar{\tilde{\zeta}}_{\bar{b}}(y) ] =
- {\bar{\alpha}^{\prime\prime}\over 2}
\sigma^{\mu}_{a\bar{b}}~\partial_{\mu}D_{m}(x-y)
\label{CCR-chiral-ghost}
\eea
and the other (anti)commutators are zero. Here
$\alpha^{\prime\prime}, \beta^{\prime\prime} \in \C$
are arbitrary parameters.

Finally we note the relations:
\bea
U(x,\theta,\bar{\theta})
= \exp (\theta\sigma^{\mu}\bar{\theta} \partial_{\mu})~
[ u(x) + 2i~\bar{\theta} \bar{\zeta}(x) + \bar{\theta}^{2}~g(x)
 ]
\nonumber \\
\tilde{U}(x,\theta,\bar{\theta})
= \exp (\theta\sigma^{\mu}\bar{\theta} \partial_{\mu})~
[ \tilde{u}(x) - 2i~\bar{\theta} \bar{\tilde{\zeta}}(x)
+ \bar{\theta}^{2}~\tilde{g}(x)
 ]
\label{chiral-gh2}
\eea
which can be used to determine the commutation relations of the superfields:
\bea
~\{ U(X_{1}),
 \tilde{U}(X_{2}) \}
= \beta^{\prime\prime}~D_{+}(X_{1};X_{2})
\nonumber \\
~\{ U(X_{1}), 
\tilde{U}^{\dagger}(X_{2}) \}
= i~\alpha^{\prime\prime}~
D_{-}(X_{1};X_{2})
\label{commutators-gh}
\eea
where
$D_{\pm}$
are given by (\ref{d-pm}).

\section{The Gauge Charge and the Gauge Supermultiplet\label{gauge-charge}}

In ordinary quantum gauge theory, one gauges away the unphysical degrees
 of
freedom of a vector field
$v_{\mu}$
using ghost fields. Suppose that the vector field is of positive mass $m$;
then one enlarges the Hilbert space with three ghost fields
$u,~\tilde{u},~\phi$
such that:
\begin{itemize}
\item
All three are scalar fields;
\item
All them have the same mass $m$ as the vector field.
\item
The Hermiticity properties are;
\be
\phi^{\dagger} = \phi, \qquad u^{\dagger} = u, \qquad
\tilde{u}^{\dagger} = - \tilde{u}
\ee
\item
The first two ones
$u, \quad \tilde{u}$
are Fermionic and
$\phi$
is Bosonic.
\item
The commutation relations are:
\bea
~[ \phi(x), \phi(y) ] = - i~D_{m}(x - y), \qquad
~\{ u(x), \tilde{u}(y) \} = - i~D_{m}(x - y)
\eea
and the rest of the (anti)commutators are zero.
\end{itemize}

Then one introduces the {\it gauge charge} $Q$ according to:

\bea
Q \Omega = 0, \qquad Q^{\dagger} = Q,
\nonumber \\
~[ Q, v_{\mu} ] = i \partial_{\mu}u, \qquad [ Q, \phi ] = i~m~u
\nonumber \\
~\{ Q, u \} = 0, \qquad
\{ Q, \tilde{u} \} = - i~(\partial^{\mu}v_{\mu} + m~\phi).
\label{gh-charge}
\eea

It can be proved that this gauge charge is well defined by these relations
i.e. it is compatible with the (anti)commutation relations.
Moreover one has
$Q^{2} = 0$
so the factor space
$Ker(Q)/Im(Q)$
makes sense; it can be proved that this is the physical space of an ensemble
of identical particles of spin $1$. For details see \cite{Sc}, \cite{Gr1}.

In \cite{GS1} we have generalised this structure for a new vector multiplet
corresponding to the representation
$\Omega_{1}$.
We try to do the same thing here for the standard vector multiplet analysed in
detail in Section \ref{vector}. 
First, it is natural to expect that the Hilbert
space
 of the model should be enlarged as above, containing beside the vector
multiplet $V$ a pair of ghost and antighost multiplets
$U, \tilde{U}$
and a scalar ghost multiplet
$H$.
The definition of the gauge charge $Q$ have to verify the consistency
relations from Section \ref{qsr}. These relations can be written in a compact
way using superfields; the non-trivial ones are

- from (\ref{susy+CCR-Q}):
\be
~[ V(X_{1}), \{ U(X_{2}), Q \} ] =-  \{ U(X_{2}) , [Q, V(X_{1}) ] \}
\label{v+u}
\ee
\be
~[ V(X_{1}), \{ \tilde{U}(X_{2}), Q \} ]  =
- \{ \tilde{U}(X_{2}) , [Q, V(X_{1}) ] \}
\label{v+tilde-u}
\ee
\be
~[ H(X_{1}), \{ U(X_{2}), Q \} ] = - \{ U(X_{2}) , [Q, H(X_{1}) ] \}
\label{h+u}
\ee
\be
~[ H(X_{1}), \{ \tilde{U}(X_{2}), Q \} ]  =
- \{ \tilde{U}(X_{2}) , [Q, H(X_{1}) ] \};
\label{h+tilde-u}
\ee

- from (\ref{susy+gauge1}):
\bea
~\left\{ Q_{a} , [ Q, V ] \right\} = - \left\{ Q , [ Q_{a}, V ] \right\}
\nonumber \\
~\left\{ Q_{a} , [ Q, H ] \right\} = - \left\{ Q , [ Q_{a}, H ] \right\}
\nonumber \\
~\left[ Q_{a} , \{ Q, U \} \right] = - \left[ Q, \{ Q_{a}, U \} \right]
\nonumber \\
~\left[ Q_{a} , \{ Q, \tilde{U} \} \right] =
- \left[ Q, \{ Q_{a}, \tilde{U} \} \right];
\label{qq1}
\eea

- from (\ref{q2}):
\bea
~\{ Q, [Q , V ] \} = 0, \qquad ~\{ Q, [ Q , H ] \} = 0,
\nonumber \\
~[ Q, \{ Q, U \} ] = 0, \qquad [ Q, \{ Q, \tilde{U} \} ] = 0.
\label{qq2}
\eea

(One has to add, of course the relations where some of the superfields are
replaced by their hermitian conjugate). Let us try to {\it define} the
gauge charge $Q$ postulating
\be
Q \Omega = 0, \qquad Q^{\dagger} = Q
\label{gauge1}
\ee
and
\be
\nonumber \\
~[ Q, V ] = U - U^{\dagger} \qquad
~\{ Q, U \} = 0.
\label{gauge2}
\ee

The action of the gauge charge on $V$ is natural if we take into account the
discussion following relation (\ref{d-prime}); it is also consistent with the
self-adjointness postulated above. Moreover it is in accordance
with the usual formul\ae~\cite{WB}. In our context it is important that
the relation ({\ref{v+u}) and the relevant relations (\ref{qq1}) and
(\ref{qq2}) are identically verified.

It will be useful to translate (\ref{gauge2}) in terms of the component
fields of the multiplet. It is easy to obtain
\bea
~[ Q, v_{\mu} ] = i \partial_{\mu}( u + u^{\dagger} )
\nonumber \\
~[ Q, C ] = u - u^{\dagger} \qquad [ Q, \phi ] = - g^{\dagger} \qquad
[ Q, d ] = {m^{2}\over 4} ( u
 - u^{\dagger} )
\nonumber \\
~\{ Q, \chi_{a} \} = 2i~\zeta_{a} \qquad
\{ Q, \lambda_{a} \}
= \sigma^{\mu}_{a\bar{b}} \partial_{\mu}\bar{\zeta}^{\bar{b}}
\nonumber \\
~\{ Q, u \} = 0 \qquad
\{ Q, g \} = 0 \qquad [ Q, \zeta_{a} ] = 0.
\label{gauge2a}
\eea

If we consider only the fields
$v_{\mu}$
and
$u$
then we are back in the framework of (\ref{gh-charge}) with the substitution
$u \rightarrow u + u^{\dagger}$

We now need the action of the gauge charge on the antighost superfield. It is
easier to express the consistency relation (\ref{v+tilde-u}) in component
fields. We have:
\bea
~[ C(x_{1}), \{ \tilde{u}(x_{2}), Q \} ]  =
- \{ \tilde{u}(x_{2}) , [Q, C(x_{1}) ] \}
\nonumber \\
~[ d(x_{1}), \{ \tilde{u}(x_{2}), Q \} ]  =
- \{ \tilde{u}(x_{2}) , [Q, d(x_{1}) ] \}
\nonumber \\
~[ \phi(x_{1}), \{ \tilde{u}(x_{2}), Q \} ]  =
- \{ \tilde{u}(x_{2}) , [Q, \phi(x_{1}) ] \}
\nonumber \\
~[ \phi^{\dagger}(x_{1}), \{ \tilde{u}(x_{2}), Q \} ]  =
- \{ \tilde{u}(x_{2}) , [Q, \phi^{\dagger}(x_{1}) ] \}
\nonumber \\
~[ v_{\mu}(x_{1}), \{ \tilde{u}(x_{2}), Q \} ]  =
- \{ \tilde{u}(x_{2}) , [Q, v_{\mu}(x_{1}) ] \}.
\label{vut}
\eea

Because
$\tilde{u}$
is a Fermi field, the expression
$\{Q, \tilde{u} \}$
must be a Bose field. From Poincar\'e covariance arguments we have the ansatz
\bea
\{ Q, \tilde{u} \} = \delta_{1}~C + \delta_{2}~d + \delta_{3}~\phi
+ \delta_{4}~\phi^{\dagger} + \delta_{5}~\partial^{\mu}v_{\mu} - i~h
\label{gauge3}
\eea
where
$\delta_{j}, j = 1,\dots,5$
are some complex number and $h$ is a complex scalar field, such that
$H = s(h)$.
We will prove later that one cannot take
$h = 0$.
If we substitute (\ref{gauge3}) in (\ref{vut}) and use (\ref{CCR-V}) and
(\ref{CCR-chiral-ghost}) we get the following system of equations for the
numbers
$\delta_{j}$:
\bea
\delta_{1} + \alpha~\delta_{2} + i~\beta~\delta_{3} - i~\bar{\beta}~\delta_{4}
= \bar{\alpha^{\prime\prime}}
\nonumber \\
\alpha~\delta_{1} + {m^{4}\over 16}~\delta_{2}
+ {i~m^{2}~\beta\over 4}~\delta_{3} - {i~m^{2}\bar{\beta}\over 4}~\delta_{4}
= {m^{2}\over2} \bar{\alpha^{\prime\prime}}
\nonumber \\
\beta~\delta_{1} + {m^{2}\beta\over 4}~\delta_{2}
- i\left( {~m^{2}\over 4} + \alpha \right)~\delta_{4}
+ i~m^{2}\beta~\delta_{5} = 0
\nonumber \\
\bar{\beta}~\delta_{1} + {m^{2}\bar{\beta}\over 4}~\delta_{2}
+ i\left( {~m^{2}\over 4} + \alpha \right)~\delta_{3}
- i~m^{2}\bar{\beta}~\delta_{5} = \beta^{\prime\prime}
\nonumber \\
\beta~\delta_{3} + \bar{\beta}~\delta_{4}
- 2~\left( {~m^{2}\over 4} + \alpha \right)~\delta_{5} =
- i~\bar{\alpha^{\prime\prime}}.
\label{system}
\eea

Now we apply the operator $s$ (the supersymmetric extension) to the relation
(\ref{gauge3}) and obtain
\be
\{ Q, \tilde{U} \} = \delta_{1}~V + \delta_{2}~D + \delta_{3}~\Phi
+ \delta_{4}~\Phi^{\dagger} + \delta_{5}~\partial^{\mu}V_{\mu} - i~H
\ee
or, if we use (\ref{algebra})
\be
\{ Q, \tilde{U} \} = \lambda_{1}~V + \lambda_{2}~{\cal D}^{2}V
+ \lambda_{3}~\bar{\cal D}^{2}V
+ \lambda_{4}~{\cal D}^{2}~\bar{\cal D}^{2}V
+ \lambda_{5}~\bar{\cal D}^{2}~{\cal D}^{2}V - i~H
\label{gauge3a}
\ee
where
\be
\lambda_{1} = \delta_{1} - {m^{2}\over 4} \delta_{2} \quad
\lambda_{2} = - {1\over 4} \delta_{3} \quad
\lambda_{3} = - {1\over 4} \delta_{4} \quad
\lambda_{4} = {1\over 32} ( \delta_{2} - 2i~\delta_{5}) \quad
\lambda_{5} = {1\over 32} ( \delta_{2} + 2i~\delta_{5} ).
\ee

One can obtain the expressions of these parameters in a different way
using directly the relations (\ref{v+tilde-u}) and the relations derived at
proposition \ref{dd}.

Let us apply the operator
${\cal D}_{a}$
to the relation (\ref{gauge3a}) and take into account that
$\tilde{U}$
is a chiral superfield. We get
\be
\lambda_{1}~{\cal D}_{a}V + \lambda_{3}~{\cal D}_{a}\bar{\cal D}^{2}V
+ \lambda_{5}~{\cal D}_{a}\bar{\cal D}^{2}~{\cal D}^{2}V - i~{\cal D}_{a}H = 0
\ee
and it follows that we should have
\be
\lambda_{j} = 0, \quad j = 1,3,5
\ee
and $H$ must be a chiral superfield. If we redefine
$
\lambda \equiv \lambda_{4}, \lambda^{\prime} \equiv \lambda_{2}
$
then we obtain
\be
\{ Q, \tilde{U} \} = \lambda~{\cal D}^{2}~\bar{\cal D}^{2}V
+ \lambda^{\prime}~{\cal D}^{2}V - i~H.
\label{gauge4}
\ee

Moreover, the system (\ref{system}) reduces to two equations which can
be taken as the definition of the parameters
$\alpha^{\prime}$
and
$\beta^{\prime}$
appearing in the causal anti-commutation relations of the ghost and antighost
superfields (\ref{CCR-chiral-ghost}) or (\ref{commutators-gh}):
\bea
\alpha^{\prime\prime} =
4 ( 4\alpha + m^{2} ) \lambda + 4i~\beta~\bar{\lambda}^{\prime}
\nonumber \\
\beta^{\prime\prime} =
16 m^{2} \bar{\beta}~\lambda - i (4 \alpha + m^{2} )~\lambda^{\prime}.
\label{ab1}
\eea

The presence of the superfield $H$ makes possible the fulfilment of all
conditions listed above. Indeed, let us we determine the action of the gauge
charge on the chiral superfield $H$. This follows from the last relation
(\ref{qq2}):
\be
~[ Q, H] = - 16 i~ m^{2}~\lambda U + i~\lambda^{\prime} {\cal D}^{2}U^{\dagger};
\label{gauge5}
\ee
in particular, this shows that we cannot have
$H = 0$.

Next, one can compute the causal commutation relation for the superfield $H$.
Using (\ref{h+tilde-u}) and the relation with
$H \rightarrow H^{\dagger}$
one can obtain relations of the type (\ref{commutators}) with
\bea
\alpha^{\prime} = 4 [ (4 \alpha + m^{2} ) (4 m^{2} |\lambda|^{2} +
|\lambda^{\prime}|^{2} ) + 32 m^{2} \lambda~Im(\beta \lambda^{\prime}) ]
\nonumber \\
\beta^{\prime} =
16 [ 16 m^{4} \bar{\beta}~\lambda^{2} - 4 \beta~(\lambda^{\prime})^{2}
- i \lambda\lambda^{\prime} (4 \alpha + m^{2} ) ]
\label{ab2}
\eea
so we have the supplementary condition
$\alpha^{\prime} > 0$.
However, one can get rid of the parameter
$\lambda$
if one performs the rescalings
\be
\tilde{U} \rightarrow - 16 \lambda \tilde{U} \qquad
H \rightarrow - 16 m~H;
\ee
In this way the preceding relations are
\be
\{ Q, \tilde{U} \} = - {1\over 16}~{\cal D}^{2}~\bar{\cal D}^{2}V
+ \lambda^{\prime}~{\cal D}^{2}V - i~m~H
\label{gauge4a}
\ee
and
\be
~[ Q, H] = i~m~U - {i\over m}~\lambda^{\prime} {\cal D}^{2}U^{\dagger}
\label{gauge5a}
\ee
and the parameters
$\alpha^{\prime}, \beta^{\prime}$
are rescaled by a factor
${1 \over m^{2}}$.
The parameters
$\alpha, \beta, \lambda^{\prime}$
remain arbitrary and all other relations of consistency are valid so
the gauge structure of the (quantum) vector field is completely determined.

If we want to have complete analogy to the usual ghost structure associated
to a massive vector field (see for instance \cite{Sc}, \cite{Gr1}) we get
new conditions on the parameter
$\lambda^{\prime}$.
Let us note from (\ref{gauge2a}) that the ghost field relevant to
$
v_{\mu}
$
is
$
u + u^{\dagger};
$
if we split $u$ into the hermitian and the anti-Hermitian part
$
u = u_{1} + i~u_{2},
$
such that
$
u_{j}^{\dagger} = u_{j};
$
then the relevant (Hermitian) ghost field is
$
2~u_{1}.
$
Suppose that we decompose
$
\tilde{u} = \tilde{u}_{1} + \tilde{u}_{2}
$
where now
$
\tilde{u}_{j}^{\dagger} = - \tilde{u}_{j};
$
then the anti-ghost field associated to
$v_{\mu}$
should be
$2~\tilde{u}_{1}$
or, up to a sign
$
\tilde{u} - \tilde{u}^{\dagger}.
$
To have complete analogy to the usual action of the gauge charge on the
antighost field we compute the expression
$
\{ Q, \tilde{u} - \tilde{u}^{\dagger} \}
$
and we get directly from (\ref{gauge3}):
\bea
\{ Q, \tilde{u} - \tilde{u}^{\dagger}\} =
(\delta_{1} - \bar{\delta_{1} })~C + (\delta_{2} - \bar{\delta_{2}})~d
+ (\delta_{3} - \bar{\delta_{4}})~\phi
+ (\delta_{4} - \bar{\delta_{3} })~\phi^{\dagger}
+ (\delta_{5} - \bar{\delta_{5}})~\partial^{\mu}v_{\mu}
\nonumber \\
- i~(h + h^{\dagger}).
\eea
This should be compared with the last relation (\ref{gh-charge}) which shows
that we must take
\be
\delta_{1} = \bar{\delta_{1}} \quad
\delta_{2} = \bar{\delta_{2}} \quad
\delta_{3} = \bar{\delta_{4}} \quad
\delta_{5} = - \bar{\delta_{5}}
\ee
and the scalar ghost field relevant to
$v_{\mu}$
should be
$h + h^{\dagger}$.

The preceding conditions give
\be
\lambda^{\prime} = 0
\ee
so the gauge structure of the vector field should be
\be
v_{\mu}, u + u^{\dagger}, \tilde{u} - \tilde{u}^{\dagger}, h + h^{\dagger}.
\label{gh-structure}
\ee

The gauge transformations of
$\tilde{U}$
and $H$ became very simple in this case:
\bea
\{ Q, \tilde{U} \} = - {1\over 16}~{\cal D}^{2}~\bar{\cal D}^{2}V - i~m~H
\nonumber \\
~[ Q, H] = i~m~U.
\label{gauge6}
\eea
In this particular case (\ref{ab2}) and (\ref{ab1}) become
\bea
\alpha^{\prime} = - (4 \alpha + m^{2} ) \qquad
\beta^{\prime} = - m^{2} \bar{\beta}
\eea
and respectively
\bea
\alpha^{\prime\prime} = - \left( \alpha + {m^{2}\over 4} \right) \qquad
\beta^{\prime\prime} = m^{2} \bar{\beta};
\eea
in particular the condition
$\alpha^{\prime} > 0$
gives
$4 \alpha + m^{2} < 0$.

The expression
$\{ Q, \tilde{U} \}$
should be compared with the expression (6.2.23) from \cite{GGRS}. The
difference is due to the fact that our definition of chirality corresponds
to the definition of anti-chirality in the standard literature and we have
$m > 0$.

\section{The Problem of Gauge Invariant Couplings\label{sm}}

To be able to construct a supersymmetric extension of a gauge model, let us
remind the reader some
 important difference between the classical and quantum
treatment of gauge
 theories. In the classical framework, we start form a Lie
algebra
$
{\mathfrak g}
$
with basis
$e_{j}, j = 1,\dots,r$
and with the Lie bracket
$[ \cdot, \cdot ]$;
the structure constants in this basis will be denoted by
$f_{jkl}$.
The basic variables of a gauge model are some classical fields
$
v_{\mu} : \R^{4} \rightarrow {\mathfrak g}
$
called the {\it gauge potentials}. We denote the set of all gauge potential
by
${\cal M}$;
its elements are also called {\it mathematical configuration}. On this set
there is an action of the gauge group, more precisely the associated gauge
algebra:
$
Gau({\mathfrak g})
$
which is by definition the set of smooth maps
$
\xi: \R^{2} \rightarrow g
$
with the pointwise Lie bracket. The action is non-linear:
\be
(\xi\cdot v)_{\mu}(x) = [ \xi(x), v_{\mu}(x) ] + \partial_{\mu}\xi(x).
\ee

By definition the {\it physical} configurations are described by the factor set
$
{\cal M}_{phys} \equiv {\cal M}/Gau({\mathfrak g}).
$
In this context the operation of {\it chosing a gauge} is perfectly
meaningful: it means to choose a section of the fibre bundle
${\cal M} \rightarrow {\cal M}_{phys}$;
every point of the section will represent a physical configuration.

If one tries to find out a (classical) Lagrangian $L$ which is invariant
with respect to this transformation, so the classical trajectories will
factorise to the set of physical configuration
$
{\cal M}_{phys}
,
$
then one essentially obtains the expression
\be
L_{YM} = <f_{\mu\nu}, f^{\mu\nu}>
\ee
where
$<\cdot,\cdot>$
is the Killing-Cartan form and
\be
f_{\mu\nu} \equiv \partial_{\mu}v_{\nu} - \partial_{\nu}v_{\mu}
+ [ v_{\mu}, v_{\nu} ]
\ee
is the well-known {\it field strength}. (If one considers Lagrangians which
are invariant  with respect to gauge transformations up to a total
divergence one also gets Chern-Simons terms). The proper mathematical
framework for this scheme is the fibre bundle theory. The inclusion of
the ghost fields  is done considering the cotangent bundle. This means
that we have to enlarge the configuration space
${\cal M}$
adding some Grassmann valued variables
$
u, \tilde{u}: \R^{4} \rightarrow {\mathfrak g} \otimes {\cal G}
$
where
${\cal G}$
is some Grassmann algebra. In this way the gauge transformation
given above is extended to the {\bf classical} BRST transformation and the
invariance of the Lagrangian with respect to the BRST transformation is
achieved by adding the Faddeev-Popov term:
\be
L_{FP} = < v_{\mu}, [ u, \partial^{\mu}\tilde{u} ] >
\ee
and a gauge fixing term
\be
L_{gf} = {1\over 2\xi} <\partial^{\mu}v_{\mu}, \partial^{\mu}v_{\mu}>.
\ee

Now, in quantum mechanics the meaning of a non-linear transformation is less
clear. However, a ``miracle" happens \cite{Sc}, \cite{Gr1}!  Let we consider
that: (a) the fields
$
v_{\mu}, u, \tilde{u}
$
are quantum free fields with the usual assignment of spin and statistics;
(b) the total Lagrangian has terms of order
$j = 2,3,4$
\be
L = L_{YM} + L_{FP} = \sum_{j=2}^{4} L^{(j)};
\ee
we promote the {\bf tri-linear} terms from the total Lagrangian to the status
of 
interaction Lagrangian, in the sense of perturbation theory by adding Wick
ordering:
\be
T(x) = :L^{(3)}:
\ee
and (c) we consider only the {\it linear} part of the BRST transformation as
a quantum operator. In this way the formul\ae~(\ref{gh-charge}) appear.
Then one can show that formula (\ref{gauge}) from the Introduction is
true for some Wick polynomial
$T^{\mu}$
and moreover, the condition of gauge invariance in the second order
generates the terms of order fourth
$L^{(4)}$
of the Lagrangian $L$. So, the condition of quantum gauge invariance generates
in a natural way the expression $L$ (up to the kinematical part which is
quadratic piece
$L^{(2)}$
of $L$;
this piece of $L$ is encoded in the structure of the Fock space).

It is natural to try the same idea in a supersymmetric context. For this we
start from the {\bf classical } supersymmetric Lagrangian. It is argued
(see \cite{GGRS} formul\ae~((6.2.12) and (6.2.20) that the corresponding terms
should have the following form. Suppose that $V, U, \tilde{U}$ has values in
${\mathfrak g}$
i.e. we have in fact $r$ vector superfields
$V_{j}$
grouped in the Lie-valued expression
$
V(X) \equiv e_{j}~V_{j}(X)
$
(we sum over the dummy indices) and similarly for $U$ and
$\tilde{U}$.
Then the classical interaction Lagrangian is taken to be the sum of
\bea
L_{YM} = - {1\over 2} \left( e^{-V}{\cal D}^{a} e^{V} \right) \bar{\cal D}^{2}
\left( e^{-V}{\cal D}^{a} e^{V} \right) + H. c.
\nonumber \\
L_{FP} = (\tilde{U} + \tilde{U}^{\dagger} )
{\cal L}_{V/2} \left[ (U + U^{\dagger})
+ \coth {\cal L}_{V/2} (U - U^{\dagger} ) \right]
\eea
and a gauge fixing term; here
${\cal L}$
is the Lie derivative. In analogy to the pure Yang-Mills case we compute
the tri-linear terms and obtain, up to a super-divergence i.e. an expression
of the type
\be
{\cal D}_{a} T^{a} + \bar{\cal D}_{a} \bar{T}^{\bar{a}}
\ee
the following interaction Lagrangian:
\be
T = \sum_{j=1}^{2} T^{(j)}
\ee
with
\bea
T^{(1)} \equiv f^{(1)}_{jkl} \left[ :V_{j} ({\cal D}^{a}V_{k})
(\bar{\cal D}^{2} {\cal D}_{a}V_{k}): - H. c. \right]
\nonumber \\
T^{(2)} \equiv f^{(2)}_{jkl} :V_{j}~(U_{k} + U_{k}^{\dagger})
(\tilde{U}_{l} + \tilde{U}_{l}^{\dagger}):
\label{lagrange}
\eea
and where
$f^{(j)}_{jkl}, j = 1,2$
are some constants proportional to the structure constants
$f_{jkl}$.
Let us note that this Lagrangian is non-renormalizable: it has the
supersymmetric canonical dimension $5$. In principle one can hope that
the gauge invariance condition will eliminate the arbitrariness
in every order of the perturbation theory such that the series of the
exponential from the classical expression
$L_{YM}$
is reconstructed perturbatively (as one get the fourth degree term of the
usual Yang-Mills Lagrangian in the  second order of the perturbation theory).
If one computes the corresponding expressions $t$ and
$t^{\mu}$
(see the Introduction) by integrating out the Grassmann variables one gets,
up to finite renormalizations, the usual expressions from the literature
\cite{Sc}, \cite{Gr1}:
\bea
\int d\theta^{2} d\bar{\theta}^{2} T^{(1)} =
4 i~f^{(1)}_{jkl} :v^{\mu}_{j} v^{\nu}_{k}~f_{l\nu\mu}: + \cdots
\nonumber \\
\int d\theta^{2} d\bar{\theta}^{2} T^{(2)} =
{i\over 2}~f^{(2)}_{jkl} :v^{\mu}_{j} (u_{k} + u_{k}^{\dagger})~
(\tilde{u}_{l} - \tilde{u}_{l}^{\dagger}): + \cdots
\eea
where by
$\cdots$
we mean terms containing the superpartners from the corresponding multiplets.
It seems incouraging that the last term is in agreement with the gauge
structure (\ref{gh-structure}). In the usual case \cite{Sc}, \cite{Gr1}
the gauge invariance is restored by adding new couplings with some scalar
ghost fields; in our case these couplings must include the Wick monomials
\bea
:(h_{j} + h_{j}^{\dagger})~(u_{k} + u_{k}^{\dagger})~
(\tilde{u}_{l} - \tilde{u}_{l}^{\dagger}):
\nonumber \\
:(h_{j} + h_{j}^{\dagger})~\partial_{\mu}(h_{k} + h_{k}^{\dagger})~v_{l}^{\mu}:
\nonumber \\
:(h_{j} + h_{j}^{\dagger})~v_{k\mu}~v_{l}^{\mu}:
\eea

Guided by this argument we study the gauge invariance of a Lagrangian of the
form
\be
T = \sum_{j=1}^{6} T^{(j)}
\ee
where the new terms are
\bea
T^{(3)} = f^{(3)}_{jkl}~:(H_{j} + H_{j}^{\dagger})~
(U_{k} - U_{k}^{\dagger}) (\tilde{U}_{l} + \tilde{U}^{\dagger}_{l}):
\nonumber \\
T^{(4)} = f^{(4)}_{jkl}~:(H_{j} + H_{j}^{\dagger})~
(H_{k} - H_{k}^{\dagger}) V_{l}:
\nonumber \\
T^{(5)} = f^{(5)}_{jkl}~:(H_{j} + H_{j}^{\dagger})~V_{k}~
\left({\cal D}^{2} \bar{\cal D}^{2}
+ \bar{\cal D}^{2} {\cal D}^{2} \right)V_{l}:
\nonumber \\
T^{(6)} = f^{(6)}_{jkl}~:(H_{j} + H_{j}^{\dagger})~V_{k}~V_{l}:
\eea
and we impose the supersymmetric gauge invariance condition (\ref{gauge-susy})
from the Introduction.

The naturalness of the new terms follows from the explicit expressions for
$d_{Q} T^{(i)}$: we have
\bea
d_{Q} T^{(1)} =
- f^{(1)}_{jkl}~:(U_{j}  + U_{j}^{\dagger})~V_{k})
({\cal D}^{2} \bar{\cal D}^{2} + \bar{\cal D}^{2} {\cal D}^{2})V_{l}:
+ 16 f^{(1)}_{jkl} m_{l}^{2} :(U_{j}  + U_{j}^{\dagger})~V_{k})~V_{l}:
\nonumber \\
+ {\cal D}_{a} T^{a} + \bar{\cal D}_{a} \bar{T}^{\bar{a}}
\eea
where
\be
T_{a} \equiv - f^{(1)}_{jkl} :(U_{j} + U_{j}^{\dagger})~V_{k}~
(\bar{\cal D}^{2} {\cal D}_{a}V_{l}):
- 2 f^{(1)}_{jkl} :U_{j}~(\bar{\cal D}_{\bar{b}}V_{k})~
{\cal D}_{a}\bar{\cal D}^{\bar{b}}V_{l}:
\ee
and the complete antisymmetry of the constants
$f^{(1)}_{jkl}$
was used. Also
\bea
d_{Q} T^{(2)} = f^{(2)}_{jkl}~:(U_{j} - U_{j}^{\dagger})~
(U_{k} + U_{k}^{\dagger})~(\tilde{U}_{l} + \tilde{U}_{l}^{\dagger})
:
\nonumber \\
+ {1\over 16}~f^{(2)}_{jkl}~:V_{j}~(U_{k} + U_{k}^{\dagger})~
({\cal D}^{2} \bar{\cal D}^{2} + \bar{\cal D}^{2} {\cal D}^{2})V_{l}:
\nonumber \\
+ i~f^{(2)}_{jkl}~m_{l}:V_{j}~(U_{k} + U_{k}^{\dagger})~
(H_{l} - H_{l}^{\dagger})
:
\eea

Then one can see that to compensate the various terms one is forced to
introduce the new terms
$T^{(j)}, j = 3,\dots,6$.
The new terms seem to be a logical choice because if we integrate out the
Grassmann variables we obtain, essentially, the usual couplings of the scalar
ghosts
 \cite{Sc}, \cite{Gr1} listed above. If one requires that the expression
$T = \sum_{j=1}^{6} T^{(j)}$
does verify the supersymmetric gauge invariance condition
(\ref{gauge-susy}) then one obtains the solution
\bea
f^{(1)}_{jkl} = 0 \qquad f^{(6)}_{jkl} = 0
\nonumber \\
f^{(2)}_{jkl} = i~m_{k}~f^{(3)}_{jkl} \qquad
f^{(5)}_{jkl} = -{1\over 16}~f^{(3)}_{jkl} \qquad
f^{(4)}_{jkl} = - i~m_{k}~f^{(3)}_{jlk}.
\label{f's}
\eea

If we compute the corresponding Lagrangian
$t(x)$
we find out a strange solution of the gauge invariance condition: there is
no pure Yang-Mills coupling but one has monomials with canonical dimension
$6$ (they are produced by
$T^{(5)}$).

The negative result which we have obtained can be traced to the gauge
structure. Indeed the cancelation of the coefficient of the Wick
monomial
$
:(U_{j} - U_{j}^{\dagger})~
(U_{k} + U_{k}^{\dagger})~(\tilde{U}_{l} + \tilde{U}_{l}^{\dagger})
:
$
in the supersymmetric gauge invariance condition (\ref{gauge-susy}) implies
the third relation from (\ref{f's}). However if we impose only (\ref{gauge})
a weaker condition follows. Indeed we have
\be
\int d\theta^{2} d\bar{\theta}^{2}~:(U_{j} - U_{j}^{\dagger})~
(U_{k} + U_{k}^{\dagger})~(\tilde{U}_{l} + \tilde{U}_{l}^{\dagger})
:
= - {1\over 4} (m_{j}^{2} + m_{l}^{2} - m_{k}^{2})
:(u_{j} + u_{j}^{\dagger})~
(u_{k} + u_{k}^{\dagger})~(\tilde{u}_{l} - \tilde{u}_{l}^{\dagger})
:
\ee
But the condition of cancelation of the coefficient of this Wick monomial
gives the weaker condition
\be
(m_{j}^{2} + m_{l}^{2} - m_{k}^{2})~
(f^{(2)}_{jkl} - i~m_{k}~f^{(3)}_{jkl}) = (j \leftrightarrow k)
\ee
because of the antisymmetry in $j$ and $k$ obtained after integrating out the
Grassmann variables. The same argument works for the annulation of the
coefficient of the Wick monomial
$
:(H_{j} + H_{j}^{\dagger})~
(U_{k} - U_{k}^{\dagger})~(H_{l} - H_{l}^{\dagger})
:
$
which gives the last relation (\ref{f's}). However, because
\be
\int d\theta^{2} d\bar{\theta}^{2}~
:(H_{j} + H_{j}^{\dagger})~
(U_{k} - U_{k}^{\dagger})~(H_{l} - H_{l}^{\dagger})
:
= - {1\over 4} (m_{k}^{2} + m_{l}^{2} - m_{j}^{2})
:(h_{j} + h_{j}^{\dagger})~
(u_{k} + u_{k}^{\dagger})~(h_{l} + h_{l}^{\dagger})
:
\ee
the condition (\ref{gauge}) gives only
\be
(m_{k}^{2} + m_{l}^{2} - m_{j}^{2})~
(f^{(4)}_{jkl} + i~m_{k}~f^{(3)}_{jlk}
) = - (j \leftrightarrow l)
\ee
because of the symmetry property in $j$ and $l$.

We have tried in vain to circumvent this no-go result taking for granted the
expressions
$T^{(j)}, j = 1, 2$
which are suggested by the existing literature. To obtain weaker conditions
from the gauge invariance condition (\ref{gauge-susy}) it seems that one is
forced to change the expression
$T^{(2)}$;
a possible choice would be
\be
T^{(2)} \equiv f^{(2)}_{jkl} :V_{j}~(U_{k} - U_{k}^{\dagger})~
(\tilde{U}_{l} - \tilde{U}_{l}^{\dagger}):
\ee
because after integrating out the Grassmann variables we again obtain the usual
expression; moreover in the expression
$
d_{Q}~T^{(2)}
$
we have now the trilinear ghost term
$
f^{(2)}_{jkl}~:(U_{j} - U_{j}^{\dagger})~
(U_{k} - U_{k}^{\dagger})~(\tilde{U}_{l} - \tilde{U}_{l}^{\dagger})
:
$
with some antisymmetry property in $j$ and $k$. However, then one is forced
to change the expression
$T^{(1)}$
too. A possible choice would be
\be
T^{(1)} \equiv f^{(1)}_{jkl} :V_{j}~V_{k}~\partial_{\mu}V^{\mu}_{l}:
\ee
Adding coupling with the scalar ghost superfields and imposing the
supersymmetric gauge invariance condition (\ref{gauge-susy}) one obtains again
after integration of the Grassmann variables a strange solution with anomalous
couplings.

We find these arguments rather convincing for a negative result. We conjecture
that one cannot find out a solution $T$ verifying the supersymmetric gauge
invariance condition (\ref{gauge-susy}) and such that after integration of the
Grassmann variables we obtain the usual Yang-Mills interaction between the
gauge Bosons and the ghost fields.

One can save the model with
$
f^{(1)}_{jkl} \not= 0
$
if one imposes only (\ref{gauge})
but in this case one can prove that one can add to $T$ many other
supersymmetric Wick monomials so the arbitrariness of the interaction
Lagrangian is rather large.
 Moreover one does not have a fully supersymmetric
gauge invariance property.

We mention in the end that one can use the
$\Omega_{1/2}$
vector mulptiplet to construct a supersymmetric extension of quantum
electrodynamics. This can be done as follows. We take two Wess-Zumino
multipltes
$
(\phi^{(j))},f^{(j)}_{a}), j = 1,2
$
verifying the relations from Subsection \ref{scalar}, in particular the
relations (\ref{susy-ch}). Then we define the left and right fields
\bea
\phi_{L} \equiv \phi^{(1)} + i \phi^{(2)} \qquad
\phi_{R} \equiv \phi^{(1)} - i \phi^{(2)}
\nonumber \\
f_{La} \equiv f^{(1)}_{a} - i f^{(2)}_{a} \qquad
f_{Ra} \equiv f^{(1)}_{a} + i f^{(2)}_{a};
\eea
the expressions
$
f_{L,R}
$
are the left and right components of the electron field. Next, we define two
chiral superfields
$
\Phi_{L} \equiv s(\phi_{L}), \Phi_{R} \equiv s(\phi_{R})
$
using the sandwich formula. If we consider now the interaction Lagrangian
\be
T = \left(\Phi_{L}^{\dagger} \Phi_{L} - \Phi_{R}^{\dagger} \Phi_{R} \right)~V
\ee
between the vector superfield $V$ and these chiral superfields, then one can
prove rather easy two facts: (i) this Lagrangian is gauge invariant in the
sense (\ref{gauge}); (ii) after integrating out  the Grassmann variables
one obtains the usual expression for the QED interaction Lagrangian
\be
\int d\theta^{2} d\bar{\theta}^{2} T = v_{\mu}
\left( f_{L} \sigma^{\mu} \bar{f}_{L} - f_{R} \sigma^{\mu} \bar{f}_{R} \right)
+ \cdots
\ee
\section{The Linear Vector Model\label{lin}}

We try to circumvent the negative result from the preceding Section
by choosing a {\it gauge}. This operation has a perfectly well meaning in
the classical field theory context: it means to choose a section of the
fibre bundle
$
{\cal M}~\rightarrow~{\cal M}_{phys}
$;
every point of the section will represent a physical configuration. In the
quantum context, the relations (\ref{gauge2a}) are considered as a proof
that by choosing conveniently the expressions
$u, g, \psi_{a}$
one can make equal to zero the fields
$C, \phi, \chi$
and the longitudinal part of
$v_{\mu}$.

In the quantum context, we proceed as follows. Guided by Proposition
\ref{decomp-v} we impose the following restriction on the vector field:
\be
{\cal D}^{2} \bar{\cal D}^{2}~V = 0;
\label{v0}
\ee
this implies that the chiral and antichiral components
$V_{1}, V_{2}$
of $V$ are zero so we have
$V = V_{0} = - {2\over m^{2}} D^{\prime}$.
We call $V$ in this case the {\it linear} vector model.
One can express the preceding condition in component fields; it is easy to get:
that the condition (\ref{v0}) is equivalent to:
\bea
\partial^{\mu}v_{\mu} = 0 \qquad d = - {m^{2}\over 2}~C \qquad \phi = 0
 \qquad
\chi_{a} = {i\over m^{2}} \sigma^{\mu}_{a\bar{b}}
\partial_{\mu}\bar{\lambda}^{\prime\bar{b}}.
\label{restr}
\eea

From these constraints it follows that
 we have
$
d^{\prime} = 2d, \quad \lambda^{\prime}_{a} = 2 \lambda_{a}.
$
This reduction of the multiplet is consistent. Indeed, the fields are now
the scalar field $d$, the transversal vector field
$v_{\mu}$
and the Dirac field
$\lambda_{a}$
and if we act with this filed on the vacuum we get the representation
$\Omega_{1/2}$.
One can easily obtain the following action of the supercharges from
(\ref{susy-action}) if we use (\ref{restr}):
\bea
~[Q_{a}, d^{\prime} ] =
- {1\over 2} \sigma^{\mu}_{a\bar{b}}\partial_{\mu}\bar{\lambda}^{\prime\bar{b}}
\quad \Leftrightarrow \quad
~[Q_{a}, d ] =
- {1\over 2} \sigma^{\mu}_{a\bar{b}}\partial_{\mu}\bar{\lambda}^{\bar{b}}
\nonumber \\
i~[Q_{a}, v^{\mu} ] = \sigma^{\rho}_{a\bar{b}}
\left( \delta^{\mu}_{\rho} + {1\over m^{2}} \partial^{\mu}\partial_{\rho}
\right)\bar{\lambda}^{\prime\bar{b}}
= 2~\sigma^{\rho}_{a\bar{b}}
\left( \delta^{\mu}_{\rho} + {1\over m^{2}} \partial^{\mu}\partial_{\rho}
\right)\bar{\lambda}^{\bar{b}}
\nonumber \\
~\{ Q_{a}, \lambda^{\prime}_{b} \} =
2i~\epsilon_{ab} d^{\prime} - 2i~\sigma^{\mu\rho}_{ab} \partial_{\mu}v_{\rho}
\quad \Leftrightarrow \quad
~\{ Q_{a}, \lambda_{b} \} =
2i~\epsilon_{ab} d - i~\sigma^{\mu\rho}_{ab} \partial_{\mu}v_{\rho}
\nonumber \\
\{ Q_{a}, \bar{\lambda}^{\prime}_{\bar{b}} \} = 0 \quad \Leftrightarrow \quad
\{ Q_{a}, \bar{\lambda}_{\bar{b}} \} = 0;
\label{linear}
\eea
let us note that the second relation is consistent with the transversality
condition
$\partial^{\mu}v_{\mu} = 0$.
One can check directly that the consistency relations (\ref{susy+lorentz}
)
are true. The causal (anti)commutation relation can be obtained from
(\ref{CCR-V}) if we take into account the restrictions (\ref{restr}); we have
a solution {\it iff}
\be
\alpha = - {m^{2}\over 4} \qquad \beta = 0;
\ee
the explicit form is:
\bea
~[ d(x), d(y) ] = - {im^{4}\over 16}~D_{m}(x-y)
\nonumber \\
~[ v_{\mu}(x), v_{\rho}(y) ] =
i~\left(\partial_{\mu}\partial_{\rho} + m^{2}~g_{\mu\rho}\right)~D_{m}(x-y)
\nonumber \\
\left\{ \lambda_{a}(x), \lambda_{b}(y) \right\} = 0
\nonumber \\
\left\{ \lambda_{a}(x), \bar{\lambda}_{\bar{b}}(y) \right\} =
{m^{2}\over 4}~\sigma^{\mu}_{a\bar{b}}~\partial_{\mu}D_{m}(x-y);
\eea
let us remark that the second relations is compatible with the restriction
$\partial^{\mu}v_{\mu} = 0$.

One usually rejects this transversality condition because then the perturbative
theory of the vector field
$v_{\mu}$
is non-renormalizable: the causal distribution of
$v_{\mu}$
has order of singularity $0$ instead of $-2$. In the supersymmetric context
the situation is 
different. Indeed, the causal (anti)commutation given above
do not change
 the super-order of singularity of the causal distribution; we
still have
$
\omega(D(X_{1};X_{2})) = -2
$
so one can try to build a perturbation theory for the transversal vector model.
One does not need in this case the ghost fields so one can build the
perturbation theory starting from the Lagrangian
$T = T^{(1)}$
from (\ref{lagrange}
).
However, this Lagrangian in non-renormalizable even in the supersymmetric
context and there is no symmetry requirement which could restrict the
arbitrariness in higher orders of perturbation theory.
 Moreover, by
integrating out the Grassmann variables one gets an expression for the
interaction Lagrangian which is different from the standard expression
from the literature: no ghost fields are needed.

\newpage
\section{Conclusions}

We have succeded to construct in a rigorous way the {\bf quantum}
supersymmetric vector multiplet. Some differences from the literature already
appear: the mass of the multiplet should be necessarily strictly positive
and the Feynman propagator is different. The gauge structure associated to
this multiplet is, essentially the same as in the standard literature;
scalar superghosts have to be included because the mass is strictly positive.
However, the expression of the interaction Lagrangian suggested by the
literature does not verify a {\bf supersymmetric} gauge invariance condition
(\ref{gauge-susy}) as it is suggested by the BRST invariance of the classical
action. We are conjecturing that a no-go result can be obtained if one studies
systematically all possible Wick monomials (as it is done for the ordinary
gauge models in \cite{Sc}, \cite{Gr1}). The ways out of this negative result
are: (a) to change the gauge structure from Section \ref{gauge-charge};
(b) to abandon (\ref{gauge-susy}) and replace it by the weaker condition
(\ref{gauge}); (c) to relax the conditions (\ref{SUSY})
. The second possibility
means to accept a model which is not gauge invariant in a supersymmetric sense
(and this is the origin of the appearence of many free paramters). The first
possibility cannot be ruled out but we did not suced to find a natural
replacement of the gauge structure contained in the formul\ae~(\ref{gauge1}),
(\ref{gauge2}) and (\ref{gauge6}). The last possibility is suggested by the
analysis of the model as a classical field theory: one requires that
(\ref{SUSY}) are valid only ``up to gauge transformations". We did not
succed to give a rigorous formulation of this idea in our pure quantum context.

We emphasize again that the (new)
$\Omega_{3/2}$
vector multiplet introduced in \cite{GS1} is gauge invariant in a
supersymmetric sense and reproduces the usual Yang-Mills Lagrangian after
the integration of the Grassamnn variables.

\end{document}